\documentclass[aps, prd,twocolumn,superscriptaddress,nofootinbib,preprintnumbers]{revtex4-1}

\def\aj{AJ}%
\def\araa{ARA\&A}%
\def\apj{ApJ}%
\def\apjl{ApJ}%
\def\apjs{ApJS}%
\def\ao{Appl.~Opt.}%
\def\aap{A\&A}%
\def\mnras{MNRAS}%
\def\prd{Phys.~Rev.~D}%
\def\jcap{JCAP}%
\def\physrep{Phys.~Rep.}%
\def\procspie{Proc.~SPIE}%

\newcommand{\omm}{\Omega_{\rm m}}
\newcommand{\omb}{\Omega_{\rm b}}

\newcommand{\sqdeg}{{\rm deg}^{2}}

\newcommand{\sptp}{{\rm SPT}+\textit{Planck}}
\newcommand{\planck}{\textit{Planck}}

\newcommand{\kcmb}{\kappa_{\rm CMB}}
\newcommand{\ktsz}{\kappa_{\rm tSZ}}
\newcommand{\kcib}{\kappa_{\rm CIB}}

\newcommand{\gammat}{\gamma_{\rm t}}

\newcommand{\delg}{\delta_{\rm g}}

\newcommand{\cosmosis}{\textsc{CosmoSIS}}

\newcommand{\treecorr}{\textsc{TreeCorr}}

\newcommand{\flask}{\textsc{Flask}}

\newcommand{\redmagic}{\textsc{redMaGiC}}
\newcommand{\redmapper}{\textsc{redMaPPer}}

\usepackage{amsmath,amssymb,latexsym,times}
\usepackage{numdef}
\usepackage{slashed}
\usepackage{graphicx}
\usepackage{grffile}
\usepackage[normalem]{ulem}
\usepackage{dcolumn}
\usepackage{xcolor}
\usepackage{booktabs}
\usepackage{amssymb}
\usepackage{slashed}
\usepackage{todonotes}
\definecolor{aiiro}{HTML}{952525}
\usepackage{hyperref}
 \hypersetup{
     colorlinks=true,
     linkcolor=aiiro,
     citecolor = aiiro,      
     urlcolor=aiiro,
     }

\usepackage{dcolumn}
\usepackage{xcolor}
\usepackage{booktabs}
\usepackage{amsmath}
\usepackage{amssymb}
\usepackage{slashed}
\usepackage{todonotes}
\usepackage{multirow}
\usepackage{hyperref}
\usepackage[T1]{fontenc}
\usepackage{ae,aecompl}

\usepackage{lineno}

\usepackage{eso-pic}

\usepackage{txfonts}

\newcommand{\be}{\begin{equation}}
\newcommand{\ee}{\end{equation}}
\newcommand{\ba}{\begin{eqnarray}}
\newcommand{\ea}{\end{eqnarray}}

\newcommand{\nside}{\ifmmode N_{\mathrm{side}}\else $N_{\mathrm{side}}$\fi}
\newcommand{\npix}{\ifmmode n_{\mathrm{pix}}\else $n_{\mathrm{pix}}$\fi}
\newcommand{\Npix}{\ifmmode N_{\mathrm{pix}}\else $n_{\mathrm{pix}}$\fi}
\newcommand{\lmin}{\ifmmode \ell_{\mathrm{min}}\else $\ell_{\mathrm{min}}$\fi}
\newcommand{\lmax}{\ifmmode \ell_{\mathrm{max}}\else $\ell_{\mathrm{max}}$\fi}

\newcommand{\WMAP}{{\slshape WMAP~}}
\newcommand{\LCDM}{$ \Lambda $CDM~}

\newcommand\Tstrut{\rule{0pt}{2.6ex}}         
\newcommand\Bstrut{\rule[-0.9ex]{0pt}{0pt}}   

\defcitealias{Giannantonio2016}{G16}

\begin{document}

\title[CMB lensing tomography]
{Dark Energy Survey Year 1 Results: tomographic cross-correlations between DES galaxies and CMB lensing from SPT+{\it Planck}}


\author{Y.~Omori}
\affiliation{Department of Physics, Stanford University, 382 Via Pueblo Mall, Stanford, CA 94305, USA}
\affiliation{Kavli Institute for Particle Astrophysics \& Cosmology, P. O. Box 2450, Stanford University, Stanford, CA 94305, USA}
\affiliation{Department of Physics and McGill Space Institute, McGill University, Montreal, Quebec H3A 2T8, Canada}
\author{T.~Giannantonio}
\affiliation{Institute of Astronomy, University of Cambridge, Madingley Road, Cambridge CB3 0HA, UK}
\affiliation{Kavli Institute for Cosmology, University of Cambridge, Madingley Road, Cambridge CB3 0HA, UK}
\affiliation{Universit\"ats-Sternwarte, Fakult\"at f\"ur Physik, Ludwig-Maximilians Universit\"at M\"unchen, Scheinerstr. 1, 81679 M\"unchen, Germany}
\author{A.~Porredon}
\affiliation{Institut d'Estudis Espacials de Catalunya (IEEC), 08193 Barcelona, Spain}
\affiliation{Institute of Space Sciences (ICE, CSIC),  Campus UAB, Carrer de Can Magrans, s/n,  08193 Barcelona, Spain}
\author{E.~J.~Baxter}
\affiliation{Department of Physics and Astronomy, University of Pennsylvania, Philadelphia, PA 19104, USA}
\author{C.~Chang}
\affiliation{Kavli Institute for Cosmological Physics, University of Chicago, Chicago, IL 60637, USA}
\author{M.~Crocce}
\affiliation{Institut d'Estudis Espacials de Catalunya (IEEC), 08193 Barcelona, Spain}
\affiliation{Institute of Space Sciences (ICE, CSIC),  Campus UAB, Carrer de Can Magrans, s/n,  08193 Barcelona, Spain}
\author{P.~Fosalba}
\affiliation{Institut d'Estudis Espacials de Catalunya (IEEC), 08193 Barcelona, Spain}
\affiliation{Institute of Space Sciences (ICE, CSIC),  Campus UAB, Carrer de Can Magrans, s/n,  08193 Barcelona, Spain}
\author{A.~Alarcon}
\affiliation{Institut d'Estudis Espacials de Catalunya (IEEC), 08193 Barcelona, Spain}
\affiliation{Institute of Space Sciences (ICE, CSIC),  Campus UAB, Carrer de Can Magrans, s/n,  08193 Barcelona, Spain}
\author{N.~Banik}
\affiliation{Fermi National Accelerator Laboratory, P. O. Box 500, Batavia, IL 60510, USA}
\author{J.~Blazek}
\affiliation{Center for Cosmology and Astro-Particle Physics, The Ohio State University, Columbus, OH 43210, USA}
\affiliation{Institute of Physics, Laboratory of Astrophysics, \'Ecole Polytechnique F\'ed\'erale de Lausanne (EPFL), Observatoire de Sauverny, 1290 Versoix, Switzerland}
% ...

\collaboration{DES \& SPT Collaborations}

\date{Last updated \today}


\begin {abstract} 
We measure the cross-correlation between \redmagic$\,$galaxies selected from the Dark Energy Survey (DES) Year-1 data and gravitational lensing of the cosmic microwave background (CMB) reconstructed from South Pole Telescope (SPT) and $\planck$ data over $1289\deg^2$. When combining measurements across multiple galaxy redshift bins spanning the redshift range of $0.15<z<0.90$, 
we reject the hypothesis of no correlation at 19.9$\sigma$ significance.  When removing small-scale data points where thermal Sunyaev-Zel'dovich signal and nonlinear galaxy bias could potentially bias our results, the detection significance is reduced to 9.9$\sigma$. We perform a joint analysis of galaxy-CMB lensing cross-correlations and galaxy clustering to constrain cosmology, finding $\Omega_{\rm m} = 0.276^{+0.029}_{-0.030}$ and $S_{8}=\sigma_{8}\sqrt{\mathstrut \Omega_{\rm m}/0.3} = 0.800^{+0.090}_{-0.094}$.  We also perform two alternate analyses aimed at constraining only the growth rate of cosmic structure as a function of redshift, finding consistency with predictions from the concordance $\Lambda$CDM model.  The measurements presented here are part of a joint cosmological analysis that combines galaxy clustering, galaxy lensing and CMB lensing using data from DES, SPT and $\planck$.
\end {abstract}

\preprint{DES-2018-0347}
\preprint{FERMILAB-PUB-18-514-AE}

 \maketitle

\section {Introduction} \label{sec:intro}
 
The cosmic microwave background (CMB) is one of the fundamental cosmological observables.  Measurements of primary anisotropies in the CMB, sourced by fluctuations in the photon-baryon fluid at the time of recombination, have been used to place tight constraints on the physical properties of the Universe, most recently by the $\planck$ satellite mission \citep{Planck2018vi}. 

In addition to the information contained in the primary CMB anisotropy, there is also a wealth of information in secondary anisotropies resulting from perturbations to CMB light after the time of recombination \citep{Aghanim2008}.  A particularly interesting source of secondary anisotropy is gravitational lensing, which causes the paths of photons from the last-scattering surface to be perturbed by the matter in the Universe (see the review by \cite{Lewis2006}). These deflections, on the order of a few arcminutes \citep{Cole1989}, alter the CMB primary anisotropies by redistributing power across different angular scales and producing a non-Gaussian component to the primordial distribution of temperature anisotropies. Measurement of this non-Gaussian structure can be used to infer the total amount of deflection that has occurred in a given direction \citep{Okamoto2003,Hirata2003}.  High signal-to-noise measurements of CMB lensing have been obtained by several collaborations, including the Atacama Cosmology Telescope \citep[ACT,][]{Das2014, Sherwin:2017}, $\planck$ \citep{Planck2013xvii,Planck2015xv,Planck2018viii}, POLARBEAR \citep{Polarbear2014}, and the South Pole Telescope \citep[SPT,][]{VanEngelen2012,Story:2015,Omori2017}.   

The reconstructed CMB lensing signal is an integral of all deflections sourced by the large-scale structure between the last-scattering surface and us. Due to this projection, we cannot directly measure the evolution of structure along the line of sight by analyzing the lensing signal alone.  However, the signal from CMB lensing can be \textit{cross-correlated} with tracers of the matter distribution, such as galaxy catalogues with known redshifts. This allows us to measure the growth of structure in the Universe across cosmic time.

Galaxy density-CMB lensing cross-correlations have been detected by several groups, using a variety of data sets. The first significant detection was reported by \cite{Smith2007} correlating \WMAP data with radio galaxies from NVSS \citep{Condon1998}, which was later combined with other galaxy catalogues by \cite{Hirata2008}. Other recent galaxy-CMB lensing cross-correlation measurements include the cross-correlation with quasars \citep{Sherwin2012, Geach2013}, with optical and IR galaxies \citep{Bleem2012}, the cosmic infrared background \citep{Holder2013}, galaxy clusters \citep{Baxter2017,Planck2015XXIV}, and many others.  The first \textit{tomographic} cross-correlation analysis using multiple redshift bins from a single galaxy survey was carried out by (\cite{Giannantonio2016} , hereafter \citetalias{Giannantonio2016}) using CMB lensing data from SPT and $\planck$ and the Dark Energy Survey (DES) Science Verification (SV) galaxies.  

In this work, we update the results of \citetalias{Giannantonio2016} by measuring the cross-correlations between galaxy density from the DES Year-1 (Y1) data and a CMB lensing reconstruction using a combination of SPT and $\planck$ data in the SPT-SZ survey area.\footnote{\citetalias{Giannantonio2016} cross-correlated galaxies with CMB lensing maps from SPT and $\planck$ separately, whereas in this study, a lensing map derived from a combined temperature map presented in \cite{Chown18} is used.} The total area used in this work is nearly a factor of 10 larger than in \citetalias{Giannantonio2016}. We find a highly significant detection of the correlation between galaxy density and CMB lensing.  We subject the correlation function measurements and corresponding covariance estimates to several tests for systematic effects, finding that biases due to these systematic effects are negligible over the range of angular scales used for the main analysis.  

We use the measured galaxy-CMB lensing cross-correlations to extract cosmological information in several ways. First, assuming a fiducial cosmological model based on the results of \cite{Planck2015params}, we measure the  amplitude of our measurement relative to this model. The amplitude we obtain from this procedure can be directly compared with similar constraints obtained in previous studies. Second, we infer the linear growth function over the redshift ranges that DES is sensitive to, and compare that with the baseline $\Lambda$CDM model predictions derived from CMB observations. Two different approaches are used: (i) the $D_{\rm G}$ estimator introduced in \citetalias{Giannantonio2016}, and (ii) a method that allows us to marginalize over galaxy bias parameters and parameters associated with systematic measurement errors. Finally, we fix the lensing amplitude and growth-rate parameters to their $\Lambda{\rm CDM}$ values and simultaneously estimate cosmological and systematics parameters.

In some of these analyses, we perform joint fits to both the galaxy-CMB lensing cross-correlations and galaxy clustering measurements in order to break degeneracies with galaxy bias.

The measurements presented in this study are part of the joint analysis which also involves galaxy clustering, galaxy-galaxy lensing and cosmic shear measurements presented in \cite{DES2017cosmo} and also {galaxy weak lensing}-CMB {weak} lensing correlation presented in \cite{DESY1_shearkcmb}. The methods that will be used to combine these data sets are described in \cite{Baxter2018}, and the results are presented in \citep{DESY1_5x2}. 

This paper is structured as follows. We first review in Sec.~\ref{sec:theory} the theoretical foundations of CMB lensing and galaxy clustering; we then describe the DES, SPT and $\planck$ data we use in Sec.~\ref{sec:data}, and the analysis methods we follow in Sec.~\ref{sec:methods}. The tests for possible systematics are described in Sec.~\ref{sec:systematics} and the main results of this paper, together with their cosmological implications, are presented in Sec.~\ref{sec:results}.  We finally conclude in Sec.~\ref{sec:conclusion}.

\section {Theory} \label{sec:theory}

From the CMB lensing convergence $\kappa_{\mathrm{CMB}}$ and galaxy overdensity $\delta_{\rm g}$ fields, one can construct the auto- and cross-angular power spectra, which can be written as a function of multipole $\ell$ using the Limber approximation\footnote{See \cite{Krause2017} for a discussion regarding the validity of the Limber approximation in the DES multi-probe framework.} as:
\begin{align}
C^{\delg^{i}\delg^{i}}(\ell) &=
\int d\chi \frac{q_{\delg^i} \left(\frac{\ell + \frac{1}{2}}{\chi},\chi \right) \, q_{\delg^i} \left(\frac{\ell + \frac{1}{2}}{\chi},\chi \right) }{\chi^2} \, P_{\rm NL} \left(\frac{\ell+\frac{1}{2}}{\chi}, z(\chi)\right)\\
C^{\delg^{i}\kcmb}(\ell) &=
\int d\chi \frac{q_{\delta_{\rm g}^i} \left(\frac{\ell +\frac{1}{2}}{\chi},\chi \right) \, q_{\kappa_{\rm CMB}} (\chi) }{\chi^2} \, P_{\rm NL} \left(\frac{\ell+\frac{1}{2}}{\chi}, z(\chi)\right),
\label{eq:CNK}
\end{align}
where $\chi$ is the comoving distance to redshift $z$,  $P_{\mathrm{NL}}(k,z)$ is the non-linear matter power spectrum, and the galaxy and CMB lensing kernels are:
\begin{align}
q_{\delta_{\rm g}^i} (k,\chi) &= b_{\rm g}^{i}(k,z(\chi)) \, \frac{n_{\rm g}^i(z(\chi))}{\bar{n}_{\rm g}^i} \, \frac{dz}{d\chi} \, , \label{eq:weight_gal}\\
q_{\kappa_{\rm CMB}} (\chi) &= \frac{3H_0^2 \Omega_{\rm m}}{2c^2} \, \frac{\chi}{a(\chi)} \, \frac{\chi^* - \chi}{\chi^*} \, . 
\label{eq:weight_cmbkappa}
\end{align}
Here $b_{i}(k,z)$ is the galaxy bias, $n_{\rm g}^{i}(z)$ is the redshift distribution of the $i$-th galaxy sample with total density $\bar n_{\rm g}^i$, $a$ is the cosmological scale factor, and $\chi^{*}$ is the comoving distance to the horizon.  We adopt a linear galaxy bias model (i.e. constant value for all values of $k$)  with a single galaxy bias $b_i$ parameter for each galaxy redshift bin. Following \cite{Krause2017} and \cite{Baxter2018}, we restrict the analysis to angular scales over which the linear bias approximation is accurate. 

In order to be consistent with the filtering that has been applied to the CMB lensing maps (see \S\ref{sec:datacmb}), we multiply $C^{\delta_{\rm g}\kappa_\mathrm{CMB}}(\ell)$ by the filter function, $F(\ell)$, given by
\begin{equation}
F(\ell) = \left\{\begin{array}{lr}
        \exp (-\ell(\ell + 1)/\ell_{\rm beam}^2), & \text{for } 30 < \ell < 3000\\
        0, & \text{otherwise,}
        \end{array}\right.
\end{equation}
where $\ell_\mathrm{beam}\equiv \sqrt{16\ln 2}/\theta_\mathrm{FWHM}$, and $\theta_\mathrm{FWHM} = 5.4'$. The Gaussian filtering is equivalent to convolving the CMB lensing maps with a Gaussian beam of full width at half maximum $\theta_\mathrm{FWHM}$. 

The harmonic-space expression above can be rewritten in position space by applying a Legendre transformation, yielding the two-point correlation functions between two fields $a,b$:
\be
w^{ab} (\theta) = \sum_{\ell=0}^{\infty} \left( \frac{2 \ell + 1}{4 \pi} \right) \, P_{\ell} (\cos \theta) \, C^{ab}(\ell) F(\ell) \, ,
\label{eq:wth}
\ee
where $a,b\in\{\delg^i,\kcmb\}$, $P_{\ell}$ are the Legendre polynomials, and the summation can be truncated to $\ell_{\max} \sim 10^4$ for the angular scales of interest.

Following \cite{Krause2017} and \cite{Baxter2018}, we model potential biases in the estimation of the galaxy redshift distributions using a single additive bias parameter for each galaxy redshift bin.  The galaxy $n(z)$ is modified via
\begin{equation}
n_{{\rm g},\mathrm{unbiased}}^{i}(z)=n_{{\rm g}}^{i}(z-\Delta_{z,{\rm g}}^{i}),
\end{equation} 
where $\Delta_{z,{\rm g}}^{i}$ is the bias parameter.  The biased $n_{\rm g}^{i}(z)$ is then propagated to the $C^{\delg\kcmb}(\ell)$ as described above.

We calculate the power spectrum using the Boltzmann code CAMB\footnote{See \texttt{camb.info}.} \citep{Lewis:2000a,Howlett12} with the Halofit extension to nonlinear scales \citep{Smith:2003,Takahashi:2012} and the \cite{Bird:2002} neutrino extension. 

\section {Data} \label{sec:data}

\subsection{Galaxy Samples}
\label{sec:datagal}
Our analysis relies on data from first-year DES observations, which were taken between August 2013 and February 2014. The photometry and production of the science-grade `Gold' catalog are described in \cite{Drlica-Wagner2017}. The total footprint of Y1 observations with an exposure time $>90$ seconds per filter and a valid photometric calibration covers more than 1800 deg$^2$, which is reduced to $\sim 1500$ deg$^2$ after masking for bright stars and other poor-quality regions, including the Large Magellanic Cloud. 

For the galaxy sample in this work, we use the lens galaxy catalog from \cite{DES2017cosmo}.  The large-scale clustering properties of these galaxies are described in detail in \cite{ElvinPoole2017}.  This galaxy catalog was generated using the $\redmagic$\ algorithm \citep{Rozo2016redmagic}, which is designed to find red-sequence galaxies in photometric data. The resulting galaxy sample has a photometric redshift uncertainty of $\sigma_z/(1+z) \le  2\%$, over the entire redshift range {$0.15<z<0.90$} used in this analysis. We  split this into 5 tomographic bins of width $\Delta z=0.15$, as shown in Table~\ref{tab:samples}.

$\redmagic$\ produces luminosity-thresholded samples of constant comoving densities to avoid density variations that may lead to biases in clustering analyses. {Therefore, setting the luminosity threshold high leads to a sample with lower overall output number density.} Following \cite{ElvinPoole2017}, we {use} three different luminosity cuts ($L > 0.5 L_*$, $L > 1.0 L_*$, and $L > 1.5 L_*$); using a lower luminosity threshold for the three low redshift bins allows for a higher number density in these redshift bins while the two higher redshift bins require a higher luminosity cut to produce a sample with uniform density across the footprint.

\begin{table}
\small
\centering
\small
\centering
 \begin{tabular}{ccc}
\hline\hline
 \textsc{Luminosity}  & \textsc{Redshift range}      & $\bar{n}^{i}_{\rm g}$   \\ \hline 
 $L > 0.5 L_*$   &  0.15 -- 0.30         & 61,913             \Tstrut   \\
                     &  0.30 -- 0.45         & 158,616                      \\
                     &  0.45 -- 0.60         & 234,204                     \\
 $L > 1.0 L_*$   &  0.60 -- 0.75         & 139,707                    \\
 $L > 1.5 L_*$   &  0.75 -- 0.90       & 41,270                  \Bstrut   \\  \hline
 \textsc{Total}   &  0.15 -- 0.90       & 635,710                     \\\hline
\end{tabular}
\caption{\label{tab:samples} Summary statistics of the DES red-sequence galaxy samples (\redmagic) used throughout this paper.  The effective sky area covered by these samples is 1289 $\sqdeg$.}
\end{table}

\begin{figure}
\begin{center}
\includegraphics[width=1.00\linewidth, angle=0]{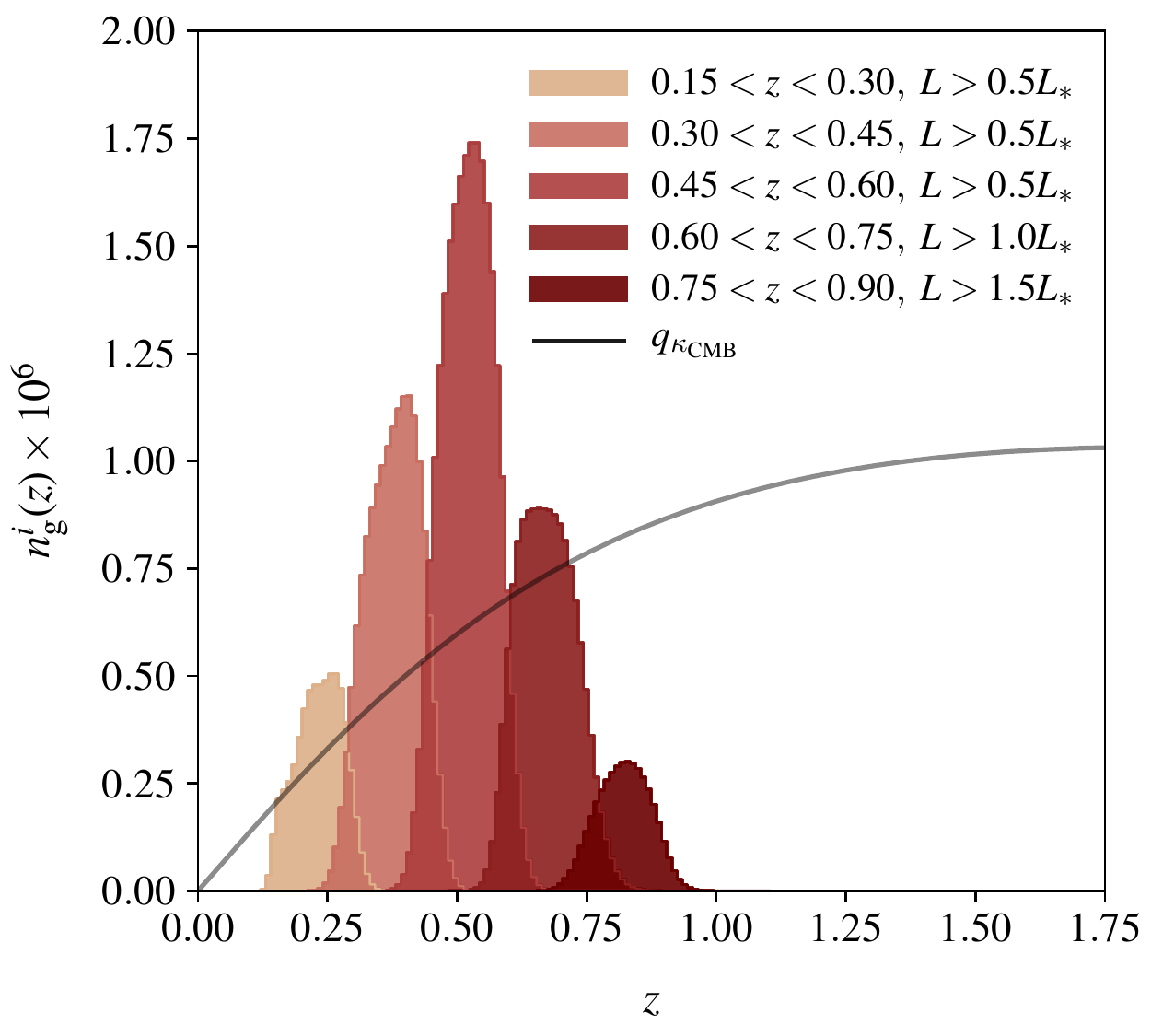}
\caption{Estimated redshift distribution of the \redmagic\ sample for the 5 tomographic bins used in this analysis. These are obtained by stacking Gaussian PDFs with mean equal to the \redmagic$\,$ redshift prediction and standard deviation equal to the resulting redshift error. Each distribution is normalized to give the total number of galaxies in each bin. {The CMB lensing efficiency $q_{\kappa_{\rm CMB}}$ is shown in black for comparison.}}
\label{fig:dndzmain}
\end{center}
\end{figure}

We estimate the redshift distributions of our galaxy samples by assuming a Gaussian redshift probability distribution function (PDF) for each object given the best-fit values for the redshift and associated error produced by the \redmagic\  algorithm. We then obtain an overall estimate of the $n^{i}_{\rm g}(z)$ of the samples by stacking these Gaussian probability distribution functions. We show the derived redshift distributions in Fig. \ref{fig:dndzmain}, from which it can be seen that the number of galaxies increases with redshift at low redshift because of the increasing volume, and decreases thereafter due to the brighter luminosity cuts imposed. 

As done in \cite{ElvinPoole2017}, the galaxy mask is constructed from the $\redmagic$ galaxy catalog over the SPT contiguous region by excluding areas outside the survey footprint, bad quality regions, and pixels with an observed coverage fraction smaller than $80\%$.  For the pixels with coverage fraction above this threshold, we assign to each mask pixel $i$ its coverage fraction $f_i$, and use this value as a weight in the clustering measurements that follow. In order to ensure the uniformity of the galaxy samples, we only use the fraction of the sky where the $L=0.5\ (1.0, 1.5)\ L_*$ galaxy sample is complete up to $z=0.6\ (0.75, 0.90)$.  We multiply this mask with the $\kcmb$ mask, which results in a combined mask with an effective area of 1289 deg$^2$.

\subsection{CMB lensing maps}
\label{sec:datacmb}

The CMB weak lensing map used in this analysis is described by \cite{Omori2017}, and we give a brief description here.  The lensing map is derived from an inverse variance weighted combination of SPT 150 GHz and Planck 143 GHz temperature data over the SPT-SZ survey region ($20^{\rm h}$ to $7^{\rm h}$  in right ascension and from $-65^\circ$ to $-40^\circ$ in declination, see, e.g. \cite{Story13}).

Modes in this combined temperature map with $\ell<100$ and $\ell>3000$ are removed to avoid foreground contamination. Point sources in the flux density range $50<F_{150}<500$ mJy  ($F_{150}>500$ mJy) in the 150 GHz band are masked with an aperture of radius $6'$ ($9'$), while sources in the flux density range $6.4<F_{150}<50$ mJy are inpainted -- the objects are masked and filled in with fluctuations similar to the CMB plus noise.  
Clusters from the SPT-detected sample of \cite{Bleem2015} with detection significance ${\rm S/N}>6$ are also masked with an aperture of $r=5'$. The quadratic estimator technique \citep{Okamoto2003} is then applied to the combined temperature map to reconstruct a filtered lensing map.  Simulations are then used to debias and renormalise the amplitude of the lensing map. In constructing the lensing map, we use the lensing multipole range of $30<\ell<3000$ and apply a Gaussian smoothing of $\theta_{\rm FWHM}=5.4'$ to the map. The low-pass filter is applied to suppress foreground contamination, while the high-pass filter is applied to remove modes we measure poorly in the data.  When calculating the correlation functions, we apply an additional stricter mask that removes all the clusters with detection significance $>5$ listed in \cite{Bleem2015} and DES $\redmapper$ clusters with richness $\lambda>80$ as well as point sources with flux density range $6.4<F_{150}<50$ mJy (which were inpainted prior to the reconstruction), in addition to that mask that was applied prior to the lensing reconstruction step.

By masking massive clusters in the CMB lensing map, we remove regions of high contamination by the tSZ effect \citep{carlstrom02}.  However, we also induce a secondary bias due to masking regions of high lensing convergence.  It was shown in \cite{Baxter2018} that this secondary bias is small compared to other systematic effects such as tSZ.  

\section {Methods} \label{sec:methods}

We measure the clustering of the galaxies and their correlations with the CMB lensing maps in position-space. Note that we take an approach slightly different from \citetalias{Giannantonio2016} here, and no smoothing is applied to the galaxy density maps. This smoothing only needs to be applied to the CMB lensing map in order to cut off the high level of noise that would otherwise leak from the high multipoles down to larger scales when transforming from harmonic to position-space.   The same smoothing was applied also to the galaxy overdensity maps in \citetalias{Giannantonio2016} for consistency, but this had the unwanted consequence of then spreading the shot-noise contribution (which in position-space is normally confined to the zero-separation bin) out to larger angular separations, thus requiring a more complex modelling (see Appendix B of \citetalias{Giannantonio2016}).

\subsection {Correlation function}

We measure both the auto-correlation of the galaxy density field and the cross-correlation between galaxies and $\kappa_{\rm CMB}$.  The former is calculated using the Landy-Szalay estimator \citep{LandySzalay93}:
\begin{align}
w^{\delg\delg} (\theta_{\alpha}) = \frac{DD(\theta_{\alpha})-2DR(\theta_{\alpha})+RR(\theta_{\alpha})}{RR(\theta_{\alpha})},
\end{align}
with
\begin{align}
DD(\theta_{\alpha})&=\frac{1}{N^{DD}_{\theta_{\alpha}}} \sum_{i = 1}^{N_{\rm gal}} \sum_{j= 1}^{N_{\rm gal}}\eta^D_i \eta^D_j \Theta_{\alpha}(\hat{\theta}^i - \hat{\theta}^j),\\
DR(\theta_{\alpha})&=\frac{1}{N^{DR}_{\theta_{\alpha}}} \sum_{i = 1}^{N_{\rm gal}} \sum_{j= 1}^{N_{\rm rand}}\eta^D_i \eta^R_j \Theta_{\alpha}(\hat{\theta}^i - \hat{\theta}^j),\\
RR(\theta_{\alpha})&=\frac{1}{N^{RR}_{\theta_{\alpha}}} \sum_{i = 1}^{N_{\rm rand}} \sum_{j= 1}^{N_{\rm rand}}\eta^R_i \eta^R_j \Theta_{\alpha}(\hat{\theta}^i - \hat{\theta}^j),
\end{align}
where $\eta^{D}$ are the weights for the individual galaxies determined from cross-correlation with systematic maps (for randoms $\eta^{R}=1$, see \cite{ElvinPoole2017} for further details), $N_{\theta}$ are the total number of pairs of a given type [data-data ($DD$), data-random ($DR$), random-random ($RR$)] in a given angular bin $\theta_{\alpha}$, and $\Theta_{\alpha}(\hat{\theta}^i - \hat{\theta}^j)$ is 1 if a pair lies at an angular distance $\theta$ within the angular bin $\alpha$ and 0 otherwise. Random galaxies are generated uniformly over the union of the galaxy and $\kcmb$ masks, and are included in the random catalog with probabilities matching the weight $f_{i}$ at the pixel which the random galaxies fall onto.\footnote{Here, we only consider the weights coming from the \emph{galaxy} mask, although both the galaxy and $\kcmb$ masks are used to determine the valid pixels.}
    
{For the correlation function between a galaxy catalog and a pixellated map such as the CMB lensing convergence map, the correlation function is calculated using:
\begin{align}
w^{\delg\kcmb} (\theta_{\alpha}) = D\kcmb(\theta_{\alpha})-R\kcmb(\theta_{\alpha}),
\end{align}
with
\begin{align}
D\kcmb(\theta_{\alpha})&=\frac{1}{N^{D\kcmb}_{\theta_{\alpha}}} \sum_{i = 1}^{N_{\rm gal}} \sum_{j= 1}^{N_{\rm pix}}
\eta^D_i \eta^{\kcmb}_j \kappa_{{\rm CMB},j}\Theta_{\alpha}(\hat{\theta}^i - \hat{\theta}^j),\\
R\kcmb(\theta_{\alpha})&=\frac{1}{N^{R\kcmb}_{\theta_{\alpha}}} \sum_{i = 1}^{N_{\rm rand}} \sum_{j= 1}^{N_{\rm pix}}
\eta^R_i \eta^{\kcmb}_j \kappa_{{\rm CMB},j}\Theta_{\alpha}(\hat{\theta}^i - \hat{\theta}^j),
\end{align}}
where $\eta^{\kcmb}_j$ is the value of the mask, and  $\kappa_{{\rm CMB},j}$ is the value of convergence at the $j$-th pixel.\footnote{Here, we only consider the weights coming from the $\kcmb$ mask, although both the galaxy and $\kcmb$ masks are used to determine the valid pixels.}  In measuring the auto-correlation of galaxy density, we use 20 bins equally spaced in logarithm between $2.5'<\theta<250'$; these angular bins are consistent with those of \cite{ElvinPoole2017}.  For  $w^{\delg\kcmb}(\theta)$, we use $10$ equally log-spaced angular bins over the same angular range due to the higher noise levels of this measurement. The measurements in both cases are carried out using the $\treecorr$ package.\footnote{\url{https://github.com/rmjarvis/TreeCorr}}

Unlike \citetalias{Giannantonio2016}, we do not perform a harmonic analysis in this paper {since the other DES-Y1 two-point analyses are all conducted in position-space, and our goal is to combine our measurements with those}. We note that $C({\ell})$ estimators allow one to get a complementary understanding of systematics to those affecting the position-space estimators and are expected to yield consistent results in terms of significance of the cross-correlation signal and corresponding cosmological implications (as discussed in detail in \citetalias{Giannantonio2016}).

\subsection{Angular scale cuts} \label{sec:scalecuts}

Our model for the correlation functions ignores several potential complications, such as the effects of tSZ bias in the CMB lensing map, the effects of non-linear galaxy bias, and the effects of baryons on the matter power spectrum.  In order to minimize biases to the inferred cosmological parameters in our analysis, we remove measurements at angular scales that we expect to be significantly impacted by these effects.  

The choices of these angular scale cuts employed here were motivated for the analyses of $w^{\delg\delg}(\theta)$ and $w^{\delg \kcmb}(\theta)$ in \cite{Krause2017} and \cite{Baxter2018}.  The scale cuts were determined by introducing unmodeled effects into simulated data vectors and performing simulated likelihood analyses to infer parameter biases.  The scale cuts ultimately chosen in \cite{Krause2017} and \cite{Baxter2018} were determined based on the joint analysis of two-point functions between the galaxy density, galaxy lensing and CMB lensing.  Since the analysis of a single correlation function --- such as $w^{\delg\kcmb }(\theta)$ --- will necessarily be less constraining, by adopting these scale cuts in this analysis we are being conservative.  It was shown in \cite{Baxter2018} that with these scale cuts, the bias on the cosmological parameter constraints will be less than $0.4\sigma$, where $\sigma$ represents the statistical uncertainty on the parameters.

The scale cut choices motivated by \cite{Krause2017} and \cite{Baxter2018} result in removing from the galaxy-CMB lensing cross-correlations angular scales that are smaller than 
\be
\theta_{\min}^{\delg\kcmb}=[15',25',25',15',15']
\ee
for the five redshift bins. The corresponding scale cuts for the galaxy auto-correlations are \citep{ElvinPoole2017}
\be
\theta_{\min}^{\delg\delg}=[45',25',25',15',15'] \, .
\ee

\subsection {Covariance matrix}

It was shown by \citetalias{Giannantonio2016} that several covariance matrix estimators (including a Gaussian analytic covariance in harmonic-space) yield consistent results for the galaxy-CMB lensing correlation. Based on this comparison and the analysis of \cite{Krause2017}, we decided to use an analytic covariance estimate described in \cite{Krause2017}, but extended to include CMB lensing cross-correlations as described by \cite{Baxter2018}. Briefly, this estimator is a sum of Gaussian covariance and non-Gaussian terms based on a halo-model approach (which includes the trispectrum term and the super-sample covariance). {We additionally modify the term of this covariance that involves correlations between $\kcmb$ noise and $\delg$ noise, to take into account the survey geometry. This is done by replacing the analytic noise-noise covariance (which is calculated based on the galaxy number density and survey area only) with the covariance calculated from correlating Gaussian realizations of the $\kcmb$ map and the galaxy random catalog using the survey mask. This correction increases the diagonal elements of the analytic covariance by $\sim10\%$ for $w^{\delg\kcmb}(\theta)$.} We compare this theoretical covariance estimate to a data-based jackknife estimate of the covariance in Appendix~\ref{sec:covmatJK}.

\subsection{Parameter constraints}

The cross-correlation between galaxy density and the CMB convergence map contains cosmological information.  To extract this information, we assume that the likelihood 
of the measured data vector $\vec{d}$ given a model $\vec{m}$, is Gaussian:
\begin{multline}
\ln \mathcal{L}(\vec{d}|\vec{m}(\vec{p}))=  -\frac{1}{2} \sum^N_{ij} \left(d_i -m_i(\vec{p})\right) \mathbf{C}^{-1}_{ij} \left(d_j - m_j(\vec{p}) \right),  
\end{multline}
where $N$ is the number of points in the data and model vectors.  The posteriors on the model parameters are given by: 
\begin{equation}
P(\vec{m}(\vec{p})|\vec{d}) \propto \mathcal{L}(\vec{d} | \vec{m}(\vec{p})) P_{\rm prior} (\vec{p}),
\end{equation}
where $P_{\rm prior}(\vec{p})$ is the prior on the model parameters.

\subsubsection{Galaxy bias and lensing amplitude constraints}

Assuming the cosmological model is tightly constrained, joint measurement of $w^{\delg\kcmb}(\theta)$ and $w^{\delg\delg}(\theta)$ allows us to simultaneously constrain galaxy bias, $b$, and an overall multiplicative bias in the $\kappa_{\rm CMB}$ map, which we call $A_{\kappa}$.  This is possible because the amplitude of the galaxy-CMB lensing cross-correlation scales with $b A_{\rm \kappa}$, while the amplitude of the galaxy clustering correlation function scales with $b^2$. 

We consider two scenarios along these lines while fixing the cosmological model to the fiducial model introduced in Sec. \ref{sec:intro}. In the first scenario we fix $A_{\kappa}=1$ and constrain the galaxy bias in each redshift bin while marginalizing over the photo-$z$ uncertainties.  The second scenario is identical to the first, but we let $A_{\kappa}$ be free.  In both cases we adopt the priors on systematics parameters presented in Table~\ref{tab:priors}. 

\subsubsection{Growth function}
\label{sec:Dg_theory}
{We use the measured correlation functions to constrain the cosmological growth function using two different methods. For both of these methods we assume $A_{\kappa }=1$}.

{The first approach is the procedure introduced in \citetalias{Giannantonio2016} (also applied in \cite{Bianchini18}).}  For this method, we fix all the cosmological and nuisance parameters to the fiducial values listed in Table \ref{tab:priors}. 
We define the growth-removed auto- and cross-spectra, indicated with a slashed symbol, via:
 \begin{align}
\slashed{C}^{\delg\delg}(\ell) &=
\int d\chi \frac{q^i_{\delta_{\rm g}} \left(\frac{\ell + \frac{1}{2}}{\chi},\chi \right) \, q^i_{\delta_{\rm g}} \left(\frac{\ell + \frac{1}{2}}{\chi},\chi \right) }{\chi^2 \, D^2[z(\chi)]} \, P_{\rm NL} \left( \frac{\ell+\frac{1}{2}}{\chi}, z(\chi)\right) \\
\slashed{C}^{\delg\kcmb}(\ell) &=
\int d\chi \frac{q^i_{\delta_{\rm g}} \left(\frac{\ell + \frac{1}{2}}{\chi},\chi \right) \, q_{\kappa_{\rm CMB}} (\chi) }{\chi^2 \, D^2[z(\chi)]} \, P_{\rm NL} \left(\frac{\ell+\frac{1}{2}}{\chi}, z(\chi)\right) ,
\label{eq:CNK_slashed}
 \end{align}
where $D(z)$ is the linear growth function. The angular power spectra are then transformed into $w (\theta)$ using Eq. \ref{eq:wth}, and our growth estimator is given by the ratio between {the} observed  and theoretical slashed correlation functions, averaged over a range of angular scales $[\theta^{\min}_{D_{\rm G}}, \theta^{\max}_{D_{\rm G}}]$:
\begin{align}\label{eq:dg}
D_{\rm G}=\left\langle \frac {w^{\delg\kcmb}_{\mathrm{observed}}(\theta)}{\slashed{w}^{\delg\kcmb}_{\mathrm{theoretical}}(\theta)} \sqrt{\mathstrut \frac{\slashed{w}^{\delg\delg}_{\mathrm{theoretical}}(\theta)}{w^{\delg\delg}_{\mathrm{observed}}(\theta)}}\right\rangle_{\theta^{\rm min}_{D_{\rm G}}<\theta<\theta^{\rm max}_{D_{\rm G}}} \, .
\end{align}
We measure this quantity for the five tomographic bins, which allows us to measure the evolution of the growth function in redshift bins (i.e. $D_{\rm G}(z_{i})$). The advantage of this estimator is that the measured quantity is  independent of galaxy bias since bias is canceled out by taking the ratio. Due to the filtering that removes $\ell<30$ in the $\kcmb$ map, the fiducial model $w^{\delg\kcmb}_{\rm theoretical}(\theta)$ reaches zero near $\theta=100'$, so we restrict our measurements to scales $\theta < \theta^{\max}_{D_{\rm G}} = 100'$.   For $\theta^{\rm min}_{D_{\rm G}}$, we conservatively choose the larger scale between the auto- and cross-correlation scale cuts of Sec.~\ref{sec:scalecuts} for each redshift bin.

In order to test for possible deviations from the baseline \LCDM model across the five redshift bins, we assume the shape of the linear growth function $D(z)$ to be fixed by the fiducial cosmology, {and we fit for a redshift-independent quantity $A_{\rm D}$ such that it minimizes:
\be
\chi^{2}=\sum_{ij}^{5}(D_{\rm obs}(z_{i})-A_{\rm D}D_{\rm fid}(z_{i}))\mathbf{C}_{ij}^{-1}(D_{\rm obs}(z_{j})-A_{\rm D}D_{\rm fid}(z_{j})),
\ee
with $D_{\rm obs}(z_{i})\equiv D_{\rm G}$ for this method. We take 50,000 multivariate Gaussian draws from the analytical covariance matrix to produce simulated $w^{\delg\delg}(\theta)$ and  $w^{\delg\kcmb}(\theta)$ data vectors, calculate $D(z)$ for each draw, and compute the covariance matrix $\mathbf{C}_{ij}$ over the ensemble of realizations.

The second method for measuring the growth function consists of simultaneously fitting $A_{\rm D}$, galaxy bias, and photo-$z$ bias to the observed auto- and cross-correlations using an Markov chain Monte Carlo (MCMC) approach. This method has an advantage of allowing us to vary over other systematic effects, such as photo-$z$ errors. For this method, we fix the cosmological parameters to the fiducial values in Table \ref{tab:priors} but vary the growth amplitude, galaxy biases and lens photo-$z$ biases over the priors given in the same table. 

\subsubsection{Cosmological parameter estimation}

Finally, we use the measurements of both $w^{\delg\kcmb}$ and $w^{\delg \delg}$ presented in this work to constrain cosmological parameters.  We generate posterior samples using the
\textsc{Multinest} algorithm \citep{Feroz2009} as implemented in the $\cosmosis$ \citep{Zuntz2015} package.
We let the photo-$z$ bias (i.e. lens $n(z)$ shift), galaxy bias and 6 cosmological parameters {($\Omega_{\rm m}$, $A_{\rm s}$, $n_{\rm s}$, $\Omega_{\rm b}$, $h$, $\Omega_{\nu}$)} vary simultaneously, {while we fix $A_{\kappa}=A_{\rm D}=1$}.  Here, $A_{\rm s}$ is the amplitude of the matter power spectrum, $n_{\rm s}$ is the spectral index, $\Omega_{\rm b}$ is the baryon density, $h$ is the unitless Hubble constant and $\Omega_{\nu}$ is the neutrino density. Priors on these parameters are summarized in Table \ref{tab:priors}. In this study we will focus on the constraints on $\Omega_{\rm m}$ and {$S_8 \equiv \sigma_{8}\sqrt{\Omega_{\rm m}/0.3}$, where $\sigma_{8}$ is the RMS amplitude of mass fluctuations on $8 h^{-1}$ Mpc scale. $S_{8}$ is defined to be approximately the most constrained cosmological parameter combination for galaxy weak lensing measurements. 

 }

\subsection{Blinding}
{This analysis was blinded throughout the study using a combination of various blinding schemes. First, the analysis pipeline was built using $\flask$ simulations and the paper was originally written assuming these data vectors. We then switched to a  scheme where we multiplied the CMB lensing map by an unknown factor in the range between 0.8 and 1.2, and shifted the cosmological parameter constraints that we obtained by an arbitrary number and removed the axes when generating figures. After the data passed all systematic checks, the measurements were repeated using a CMB lensing map without the random factor applied, and the cosmological parameter constraints were calculated without shifts.

\begin{table}
\small
\centering

\small
\centering

 \begin{tabular}{ccc}
\hline
\hline
\textsc{Parameter} & \textsc{Fiducial} &\textsc{Prior}\\\hline
\textsc{Cosmology}\\
$\Omega_{\rm m}$ & 0.309 &  [0.1, 0.9] \\ 
$A_\mathrm{s}/10^{-9}$ &$2.14$ &  [$0.5$, $5.0$]  \\ 
$n_{\rm s}$ & 0.967 & [0.87, 1.07]  \\
$w$ &  -1.0 &   \textsc{Fixed}   \\
$\omb$ &  0.0486 &  [0.03, 0.07]  \\
$h_0$  & 0.677 &  [0.55, 0.91]   \\
$\Omega_\nu h^2$ & $6.45\times 10^{-4}$ & [0.0006,0.01] \\
$\Omega_\mathrm{K}$ & $0$ & \textsc{Fixed}\\ 
$\tau$ & 0.066 & \textsc{Fixed}
\\

 \textsc{Growth Amplitude}  & \\
$A_{\rm D}$ & $1.0$ & [0.1,4.0]\\[0.15cm]

 \textsc{Lensing Amplitude}  & \\
$A_{\kappa}$ & $1.0$ & [0.1,4]\\[0.15cm]

 \textsc{Galaxy bias}  & \\
$b_{i}$ & $1.45,1.55,1.65,1.8,2.0$ & [0.8,3.0]\\[0.15cm]

 \textsc{Lens photo-$z$ error} & \\
$\Delta_{z,{\rm g}}^{1}$ & 0.010 &(0.008,0.007)\\
$\Delta_{z,{\rm g}}^{2}$ & -0.004 &
(-0.005,0.007)\\
$\Delta_{z,{\rm g}}^{3}$ & 0.009 &
(0.006,0.006) \\
$\Delta_{z,{\rm g}}^{4}$ & 0.001 &
(0.0,0.01) \\
$\Delta_{z,{\rm g}}^{5}$ & 0.0 &
(0.0,0.01) \\[0.15cm]\hline
\end{tabular}
\caption{\label{tab:priors}
The fiducial parameter values and priors for cosmological and nuisance parameters used in this analysis. Square brackets denote a flat prior over the indicated range, while parentheses denote a Gaussian prior of the form $\mathcal{N}(\mu,\sigma)$. The Gaussian priors on photo-$z$ errors are determined by \protect\cite{Cawthon2018}.
The fiducial cosmological parameter values are taken from the \cite{Planck2015params}, but here we assume 3 massive neutrinos to stay consistent with other DES-Y1 analyses. For the photo-$z$ bias, peaks of the posterior distributions in the DES joint galaxy clustering and lensing analysis \citep{DES2017cosmo} are used as fiducial values.}
\end{table}

\section {Systematic Error Analysis} \label{sec:systematics}

Systematic errors can impact the relationship between the measured and predicted correlation functions in three ways: (1) by affecting the observed density of galaxies on the sky, (2) by affecting the CMB lensing map, and (3) by affecting the inferred redshift distributions of the galaxies.  Systematics affecting the observed density of DES $\redmagic$ galaxies were explored by \cite{ElvinPoole2017} as part of the \cite{DES2017cosmo} analysis.  The main source of systematic error impacting the CMB lensing map is contamination by the tSZ effect which has been discussed and modeled in \cite{Baxter2018}.  Systematic errors in the photometric redshift distributions of $\redmagic$ galaxies were explored by \cite{Cawthon2018}, also as part of the \cite{DES2017cosmo} analysis.  Below, we draw heavily from these companion papers to constrain the systematic contamination of the measured correlation functions.

\subsection{Galaxy density and CMB lensing biases}

We first consider systematics impacting galaxy density and the CMB lensing map.  It is useful to divide these systematics into two categories: those that produce a bias that is uncorrelated with the true density fluctuations, and those that produce a bias that is correlated with them.  For those systematic biases that are uncorrelated with the true density fluctuations, in order to generate a bias in $w^{\delg\kcmb}$, the systematic must contaminate {\it both} the galaxy density and $\kcmb$; if it only impacts one of these observables, its impact on the correlation function should average to zero.  One of the strengths of cross-correlation measurements such as $w^{\delg\kcmb}$ is that there are not many systematics that could contaminate both of the observed fields.  However, there are some potential sources of bias that could do this. One example is dust, which is one of the foreground components of the CMB temperature measurements, and one can expect potential residuals in a single-frequency temperature map that can then propagate into the CMB lensing map. Dust also affects the photometry of the observed galaxies and is correlated with galactic latitude.  This contamination can then induce density fluctuations through the change of mean density with latitude.  Consequently, dust extinction may contaminate simultaneously galaxy density and CMB lensing, and could therefore contaminate measurement of $w^{\delg\kcmb}$.  In what follows we will consider dust extinction and stellar density maps as potential contaminants.

On the other hand, there are some sources of contamination which {\it are} correlated with the true density fluctuations.  In this case, the contaminant needs not affect both galaxy density and CMB lensing in order to bias $w^{\delg\kcmb}(\theta)$.  At $\sim 150$ GHz (roughly the frequencies of the SPT and $\planck$ maps used to generate the CMB lensing map), the tSZ effect results in a decrement in the observed CMB temperature around clusters. This non-Gaussian feature gets picked up as a false lensing signal by the quadratic estimator. Since hot gas is correlated with galaxies, we expect the tSZ effect to induce a bias in the measured correlation functions.  The CIB, which is dominated by emission from dusty star forming galaxies, is another extragalactic foreground that injects non-Gaussian features in our temperature maps. While the CIB emission spectrum peaks at a higher frequency, minor correlations with 150 GHz observations are expected, which again lead to false lensing signal.  Since both the CIB and tSZ originate from large-scale structure, we expect them to introduce biases in $\kappa_{\rm CMB}$ that are correlated with density fluctuations.  Maps for both tSZ and CIB contamination are built and described in detail by \cite{Baxter2018}.  
That work also identified the tSZ effect as the dominant source of systematic affecting the cross-correlation measurement between $\kappa_{\rm CMB}$ and $\delg$.   While the angular scale cuts proposed by \cite{Baxter2018} and restated in Sec.~\ref{sec:scalecuts} are chosen to mitigate these biases, they do not remove them entirely, and the residuals must be quantified. 

To quantify the impact of these potential systematics we write the observed CMB lensing signal and galaxy density fluctuations in terms of their true (cosmological) components and the \textit{known} observational or astrophysical contributions,
\begin{eqnarray}
\kappa^{\rm obs}_{\rm CMB} &=& \kappa^{\rm true}_{\rm CMB} +  \kappa_{\rm CMB}^{\mathcal{S}}, \\
\delta^{\rm obs}_{\rm g} &=& \delta^{\rm true}_{\rm g} +  \delta^{\mathcal{S}}_{\rm g},
\end{eqnarray}
where $\mathcal{S}$ refers to the systematic effects under consideration that are not expected to correlate with density fluctuations (i.e., dust and stars), and $\kappa_{\rm tSZ}$ and $\kappa_{\rm CIB}$ are the biases induced by tSZ and CIB, respectively.  We assume that $\kappa_{\rm CMB}^{\mathcal{S}}$ and $\delta_{\rm g}^{\mathcal{S}}$ are proportional to the actual spatial distribution of dust extinction and/or stars  $\mathcal{S}$, which can be estimated.  Since by definition the latter is not correlated with any of the true signals or those originated by tSZ or CIB, we have:\footnote{Here we ignore the correlation between the systematic effects.}
\begin{eqnarray}
\kappa_{\rm CMB}^{\mathcal{S}} &=& \frac{w^{\kcmb^{\rm obs} \mathcal{S}}}{w^{\mathcal{S}\mathcal{S}}}  {\mathcal{S}} \\
\delta_{\rm g}^{\mathcal{S}} &=& \frac{w^{\delg^{\rm obs}\mathcal{S}}}{w^{\mathcal{S}\mathcal{S}}}  {\mathcal{S}} \, ,
\end{eqnarray}
where $w^{\kcmb \mathcal{S}}, w^{\delg\mathcal{S}}, w^{\mathcal{S}\mathcal{S}}$ are the correlation functions between the systematic map ${\mathcal{S}}$ and the CMB lensing map, the DES galaxy density map, and with itself, respectively.  On the other hand, as described in more detail in \cite{Baxter2018}, we estimate $\kappa_{\rm tSZ}$ by passing a template tSZ map through the lensing pipeline of \cite{Omori2017}; similarly, $\kappa_{\rm CIB}$ is estimated by passing a 545\ GHz CIB map \citep{GNILC} though the \cite{Omori2017} lensing pipeline.  In total, the true cross-correlation function might receive the following unwanted contributions:
\begin{align}
w^{\delg\kcmb}_{\mathrm{obs}}(\theta) &= w^{\delg\kcmb}_{\mathrm{true}}(\theta)\nonumber\\
&+ \frac{w^{\kcmb \mathcal{S}_{\rm g}}(\theta) w^{\delg\mathcal{S}_{\rm g}}(\theta)}{w^{\mathcal{S}_{\rm g}\mathcal{S}_{\rm g}}(\theta)} + w^{\delg\kappa_{\rm tSZ}}(\theta) + w^{\delg\kappa_{\rm CIB}}(\theta) \, ,
\label{eq:syscrosseq}
\end{align}
where $S_{\rm g}$ represents dust extinction or stellar density. Using maps of these systematics, we measure the amplitude of the extra terms in the equation above, and compare them with the statistical errors on our correlation function measurements.

We first focus on those systematics that are uncorrelated with the true density.  We show in the top half of Fig.~\ref{fig:syst1} the ratio between {${w^{\kcmb \mathcal{S}}(\theta) w^{\delg\mathcal{S}}(\theta)}/{w^{\mathcal{S}\mathcal{S}}(\theta)}$  measured for} dust extinction and stellar contamination, and the uncertainty on the measured galaxy density-CMB lensing correlation
and verify that this ratio is significantly smaller than 1 and consistent with 0 across all angular scales. 
We can see that the impact on the measurements is generally small compared with the statistical error bars, so that we can conclude there is no evidence for any of these contaminants making a significant impact on our results. 

\begin{figure}
\begin{center}
\includegraphics[width=1.0\linewidth]{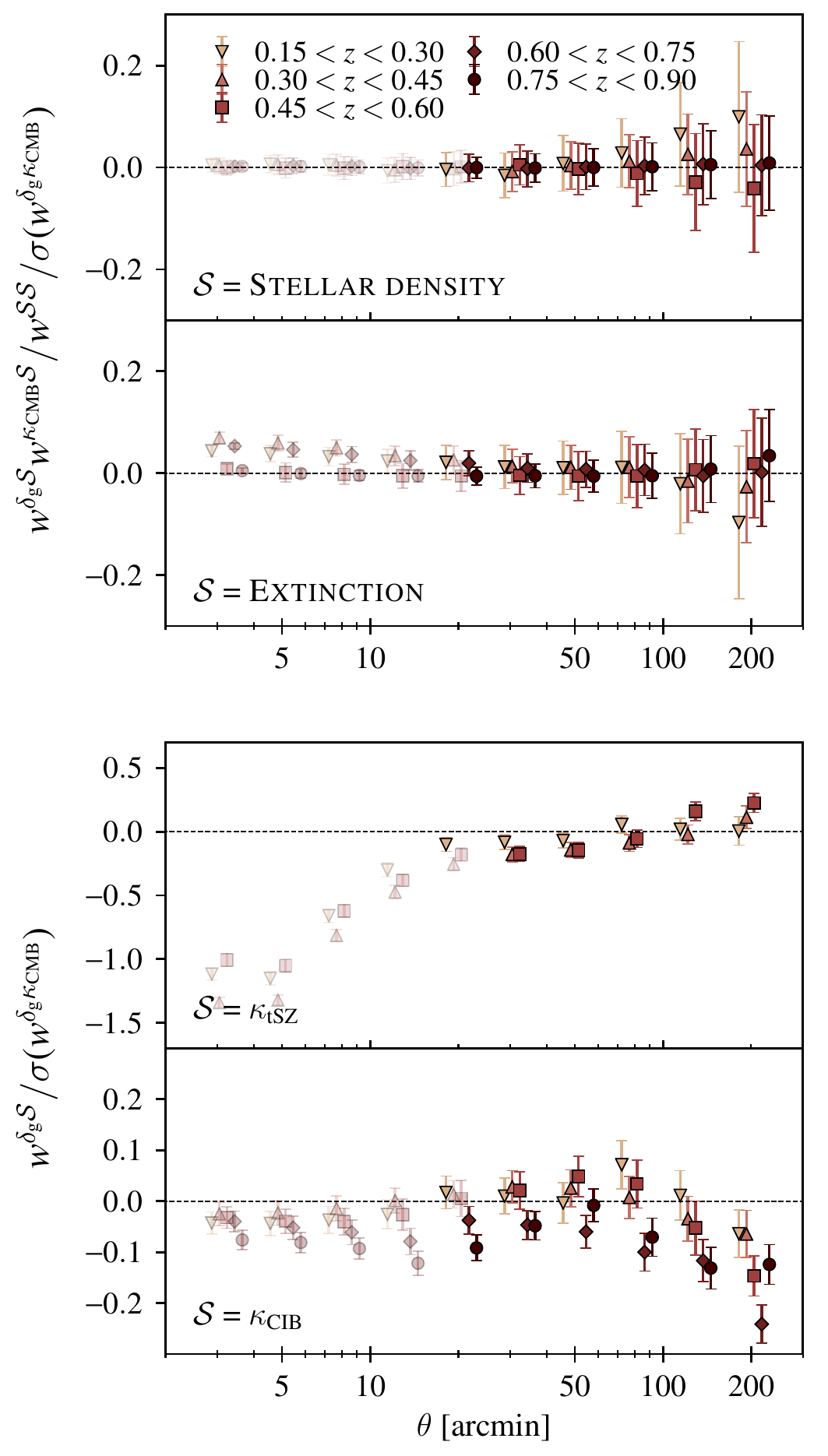}
\caption{Top two panels: Contributions due to uncorrelated systematics to the galaxy-CMB lensing cross-correlations, as described by Eq. \ref{eq:syscrosseq}, in units of the statistical errors on the observed cross-correlation.  Lower two panels: Contributions due to correlated systematics, given by the cross-correlations between the $\kcmb$ systematics ($\ktsz$ and $\kcib$) and $\delg$, also in units of the statistical error. We observe that within the angular scales we consider the ratios are $< 1$ for all redshift bins for all systematic maps (the faded points are removed from the analysis due to the imposed scale cuts). Since the tSZ template is generated only using $\redmapper$ clusters up to $z=0.6$, the correlations for the higher two redshift bins have been ignored (see \cite{Baxter2018}). {Note the different scales used for $\ktsz$ and $\kcib$.} }
\label{fig:syst1}
\end{center}
\end{figure}

We then consider the correlated sources of systematics: tSZ and CIB, and we show their contributions to Eq.~\ref{eq:syscrosseq} in the bottom half of Fig.~\ref{fig:syst1}.
Here we indeed see non-zero residuals coming from tSZ but most of this bias is removed by applying our default scale cuts, and the remaining bias is within $0.35\sigma$, where $\sigma$ is the statistical uncertainty. 

We note that \cite{ElvinPoole2017} investigated the impact of several observational systematics in addition to dust extinction and stellar density that could introduce spurious fluctuations in $\redmagic$ galaxy number density on large-scales. Using a set of 20 survey property maps,\footnote{These were exposure time, sky brightness, airmass, seeing and survey 10$\sigma$ depth, in four different broad bands.} in addition to stellar contamination and galactic extinction, they studied the dependence of number density as a function of these observational properties.  The results of these tests indicated that $\redmagic$ galaxies were not largely impacted by these systematics.  Furthermore, as mentioned above, since we do not expect the DES-specific survey systematics (exposure time, sky brightness, airmass, seeing, survey depth variations) to correlate with $\kcmb$, we do not expect these to bias $w^{\delg\kcmb}(\theta)$.

\subsection {Photo-$z$ systematics}
\label{sec:photoz}

Unlike biases in the galaxy density or CMB lensing maps, we explicitly model biases in the estimated redshift distributions of the galaxies as described in Sec.~\ref{sec:theory}.    

The \cite{DES2017cosmo} analysis constrained biases in the inferred redshift distributions of the $\redmagic$ galaxies using angular cross-correlations with spectroscopic data from the Sloan Digital Sky Survey (DR8), particularly BOSS luminous red galaxies. This analysis is presented in 
\cite{Cawthon2018}. These resultant priors on the photo-$z$ bias parameters $\Delta_{z,{\rm g}}^i$ for the five redshift bins are listed in Table \ref{tab:priors}. We let these values vary when calculating the growth amplitude and cosmological parameter constraints in Sec.\ref{sec:MCMC_Grow} and \ref{sec:MCMC_Full}.

\section {Results} \label{sec:results}

We show in Fig.~\ref{fig:wredFlask} the {measured} auto-correlation functions of the $\redmagic$ galaxy sample and its cross-correlation with the \sptp\ CMB lensing map. {The small-scale data points shown with faded symbols are the scales removed by the scale cuts as discussed in Sec.~\ref{sec:scalecuts}. The theoretical predictions assuming the fiducial cosmology listed in Table \ref{tab:priors} are shown as the black lines. In the following sections we extract cosmological information using these measurements.} 

We note that we made few modifications in Sec.~\ref{sec:directfit}
and \ref{sec:implications} after we unblinded the data.
We first added constraints on the galaxy bias from $w^{\delg\kcmb}(\theta)$ alone with the cosmology fixed to $\planck$ best-fit values (instead of DES-Y$1$). The motivation for this was to allow us to directly compare the constraints on galaxy bias when varying over the  cosmological parameters, but combining with the $\planck$ baseline likelihood. Accordingly, we also changed the assumed cosmology when computing the best-fit biases from  $w^{\delg\delg}(\theta)+w^{\delg\kcmb}(\theta)$.
Additionally, we recomputed the galaxy biases inferred from galaxy clustering and  galaxy-galaxy lensing using the same data vectors but combined with $\planck$ baseline likelihood.

\subsection{Galaxy bias and lensing amplitude} \label{sec:directfit}

\begin{figure*}
\begin{center}
\includegraphics[width=1.00\linewidth, angle=0]{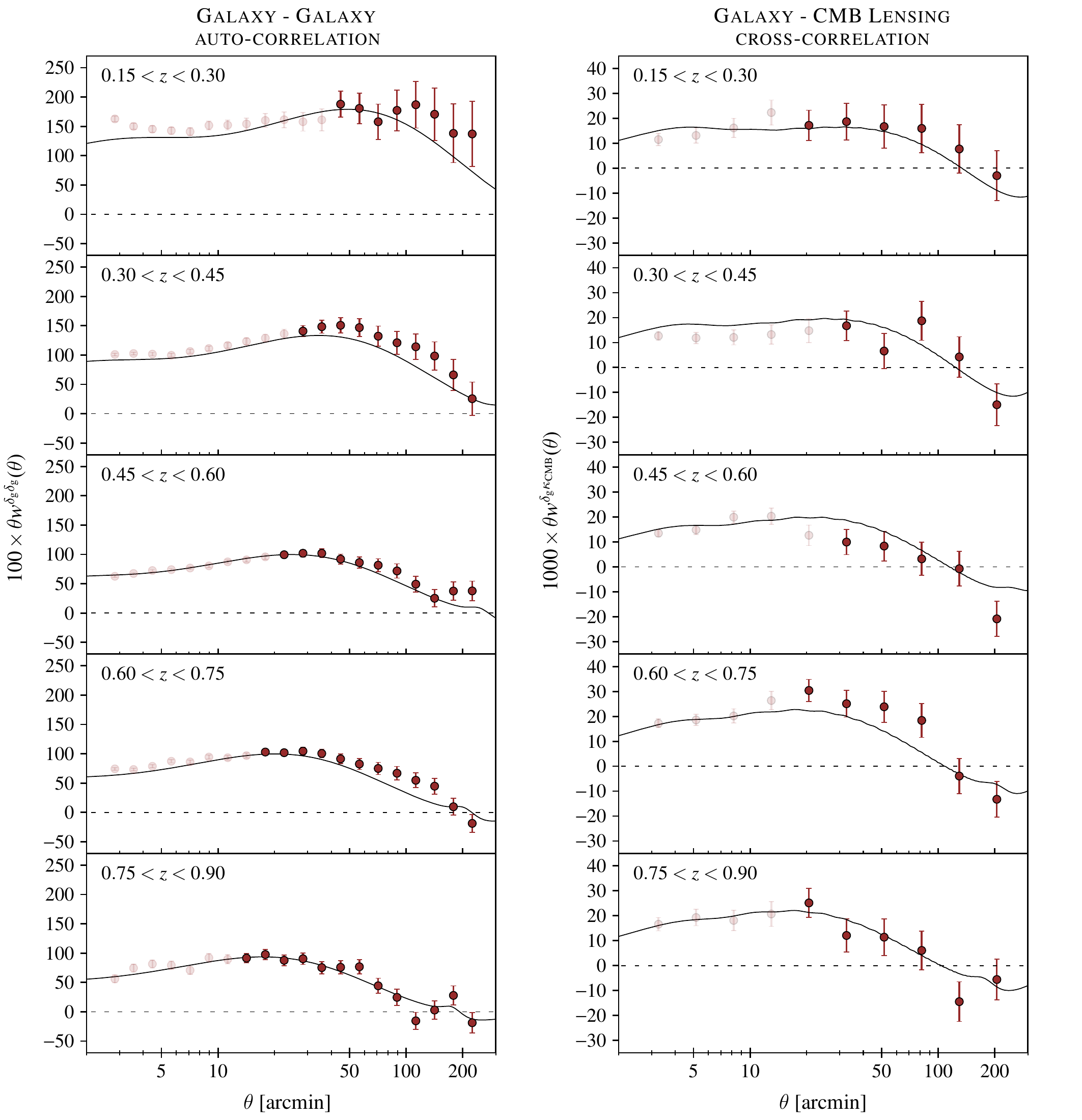}	
\caption{Measured auto- and cross-correlation functions between the $\redmagic$ galaxy sample described in Sec.~\ref{sec:datagal} and CMB lensing from \cite{Omori2017}. The faint angular bins have been excluded from the fits, consistently with \cite{ElvinPoole2017} and with \cite{Baxter2018}. The theory modeling shown uses the mean bias and cross-correlations amplitudes found in Sec.~\ref{sec:results} 
and Table~\ref{tab:fitresults}, assuming the fiducial cosmology listed in Table \ref{tab:priors}. {The error bars shown are the diagonal elements of the covariance matrix $\sqrt{\mathbf{C}_{ii}}$, and therefore, the correlations between the bins are ignored.  In contrast, the best-fit amplitudes are calculated including the off-diagonal elements and therefore the best-fit lines and data points may not match visually in certain bins. }} 
\label{fig:wredFlask}
\end{center}
\end{figure*}

We first fix the cosmological parameters to the fiducial values and vary the galaxy bias and lens photo-$z$ error parameters simultaneously, imposing the priors shown in Table~\ref{tab:priors}. We focus on constraining galaxy bias, assuming a fixed lensing amplitude of $A_{\kappa}=1$. When using both $w^{\delg\delg}(\theta)$ and $w^{\delg\kcmb}(\theta)$, we obtain $b_1 =1.48^{+0.06}_{-0.09}$, $b_2 = 1.68^{+0.05}_{-0.05}$, $b_3 =1.68^{+0.04}_{-0.04}$, $b_4 = 2.02^{+0.05}_{-0.04}$, and $b_{5} = 2.12^{+0.07}_{-0.06}$, with $\chi^2 =106.8$ for 82 data points. The high value of $\chi^2$ is primarily driven by the galaxy clustering measurements (see \cite{ElvinPoole2017}).

When we additionally treat the CMB lensing amplitude $A_{\kappa}$ as a free parameter, we obtain $b_1 =1.46^{+0.09}_{-0.07}$, $b_2 = 1.69^{+0.04}_{-0.06}$, $b_3 =1.68^{+0.05}_{-0.03}$, $b_4 = 2.04^{+0.04}_{-0.06}$, $b_{5}=2.14^{+0.05}_{-0.08}$.  The recovered posterior on the lensing amplitude is $A_{\kappa} = 1.03^{+0.13}_{-0.12}$, with a total $\chi^2 =107.2$. The similarity between the constraints on the galaxy bias values that we obtain with $A_{\kappa}$ fixed to 1 and free suggests that the galaxy bias constraints in this analysis are dominated by $w^{\delg\delg}(\theta)$.   These results are summarized in Table~\ref{tab:fitresults}.

Next, we consider constraints on galaxy bias from $w^{\delg\kcmb}(\theta)$ alone.  We reject the hypothesis of no lensing with a significance\footnote{The significance is calculated using $\sqrt{\chi^2_{\rm null}}$, where $\chi^{2}_{\rm null}$ is the value of $\chi^2$ computed under the null model, i.e. with galaxy bias $b=0$.} of $19.9 \sigma$ when no scale cuts are imposed and $9.9\sigma$ after imposing scale cuts. The constraints on galaxy bias from this analysis are summarized in Table~\ref{tab:fitresults2}.  Not surprisingly, the constraints on galaxy bias from $w^{\delg\kcmb}(\theta)$ alone are significantly weaker than in the case when the measurements are combined with $w^{\delg\delg}(\theta)$.  Similar constraints on galaxy biases are obtained when cosmological parameters are varied, but cosmological priors from \textit{Planck} baseline likelihood are imposed (right column of Table~\ref{tab:fitresults2}).

\begin{table}
\small
\centering
\small
\centering
 \begin{tabular}{ccc}
\hline
\hline
\textsc{Sample} & \textsc{Bias ($A_{\kappa}=1$)} & \textsc{Bias ($A_{\kappa}\neq1$)}  \\
\hline
\\[\dimexpr-\normalbaselineskip+3pt]
$0.15<z<0.30$ & $1.48^{+0.06}_{-0.09}$ & $1.46^{+0.09}_{-0.07}$ \\ 
$0.30<z<0.45$ & $1.68^{+0.05}_{-0.05}$ & $1.69^{+0.04}_{-0.06}$ \\ 
$0.45<z<0.60$ & $1.68^{+0.04}_{-0.04}$ & $1.68^{+0.05}_{-0.03}$\\ 
$0.60<z<0.75$ & $2.02^{+0.05}_{-0.04}$ & $2.04^{+0.04}_{-0.06}$\\ 
$0.75<z<0.90$ & $2.12^{+0.07}_{-0.06}$ & $2.14^{+0.05}_{-0.08}$\Bstrut\\[0.1cm]
\hline
\Tstrut			 & 							& $A_{\kappa} = 1.03^{+0.13}_{-0.12}$\\[0.10cm]
\hline
\end{tabular}
\caption{Summary of the constraints on the galaxy bias parameters using $w^{\delg\delg}(\theta)+w^{\delg\kcmb}(\theta)$ and assuming $\planck$ best-fit \LCDM cosmology. We consider two cases: fixing the lensing amplitude to 1 (left), and setting $A_{\kappa}$ free (right). We obtain  $\chi^2$ of 106.8 and 107.2 respectively\footnote{The fact that we obtain a marginally higher $\chi^{2}$ for the free $A_{\kappa}$ case is due to the MCMC chain not reaching the absolute maximum posterior point.}\ for 82 data points. {Since most of the constraining power is coming from $w^{\delg\delg}(\theta)$, we find similar values for the galaxy biases for the two cases. The difference between the values reported here and in \cite{ElvinPoole2017}  is due to the cosmology assumed.}  }
\label{tab:fitresults}
\end{table}

\begin{table}
\small
\centering
\small
\centering
 \begin{tabular}{ccc}
\hline
\hline
\textsc{Sample} & \textsc{Fixed} & +\textit{Planck} \textsc{Baseline}  \\
\hline
\\[\dimexpr-\normalbaselineskip+3pt]
$0.15<z<0.30$ & $1.54^{+0.44}_{-0.38}$ & $1.47^{+0.51}_{-0.38}$ \\ 
$0.30<z<0.45$ & $1.45^{+0.33}_{-0.42}$ & $1.30^{+0.46}_{-0.32}$ \\ 
$0.45<z<0.60$ & $1.10^{+0.36}_{-0.21}$ & $1.06^{+0.34}_{-0.20}$\\ 
$0.60<z<0.75$ & $2.69^{+0.23}_{-0.28}$ & $2.78^{+0.17}_{-0.35}$\\ 
$0.75<z<0.90$ & $2.17^{+0.45}_{-0.42}$ & $2.31^{+0.35}_{-0.52}$\Bstrut\\[0.1cm]
\hline
\end{tabular}
\caption{Summary of the constraints on the galaxy bias parameters from $w^{\delg\kcmb}(\theta)$ only with fixed cosmology, and in combination with the \textit{Planck} baseline likelihood. In both cases photo-$z$ biases are varied over with the priors from \cite{DES2017cosmo}. The $\chi^{2}$ that we obtain are 27.6 and 26.1 for 27 data points.}
\label{tab:fitresults2}
\end{table}

The bias constraints from $w^{\delta_{\rm g}\kcmb}(\theta)$, as well as those from the DES-Y1 galaxy clustering \citep{ElvinPoole2017} and galaxy-galaxy lensing analyses \citep{Prat2017} when combined with $\planck$ baseline likelihood are shown in Fig.~\ref{fig:bias}. We find that  considerably tighter constraints can be obtained from $w^{\delg\delg}(\theta)$ relative to $w^{\delta_{\rm g}\gammat}(\theta)$ and $w^{\delg\kcmb}(\theta)$. The constraining power of $w^{\delta_{\rm g}\kcmb}(\theta)$ relative to $w^{\delta_{\rm g}\gammat}(\theta)$ increases with higher redshift galaxy samples. This is because the number of available background galaxies decreases as we increase the redshift of the lens galaxy sample. In contrast, the signal improves for $w^{\delg\kcmb}(\theta)$ due to the better overlap with the CMB lensing kernel as shown in Fig. \ref{fig:dndzmain}.  

\begin{figure}
\begin{center}
  \includegraphics[width=1.0\linewidth, angle=0]{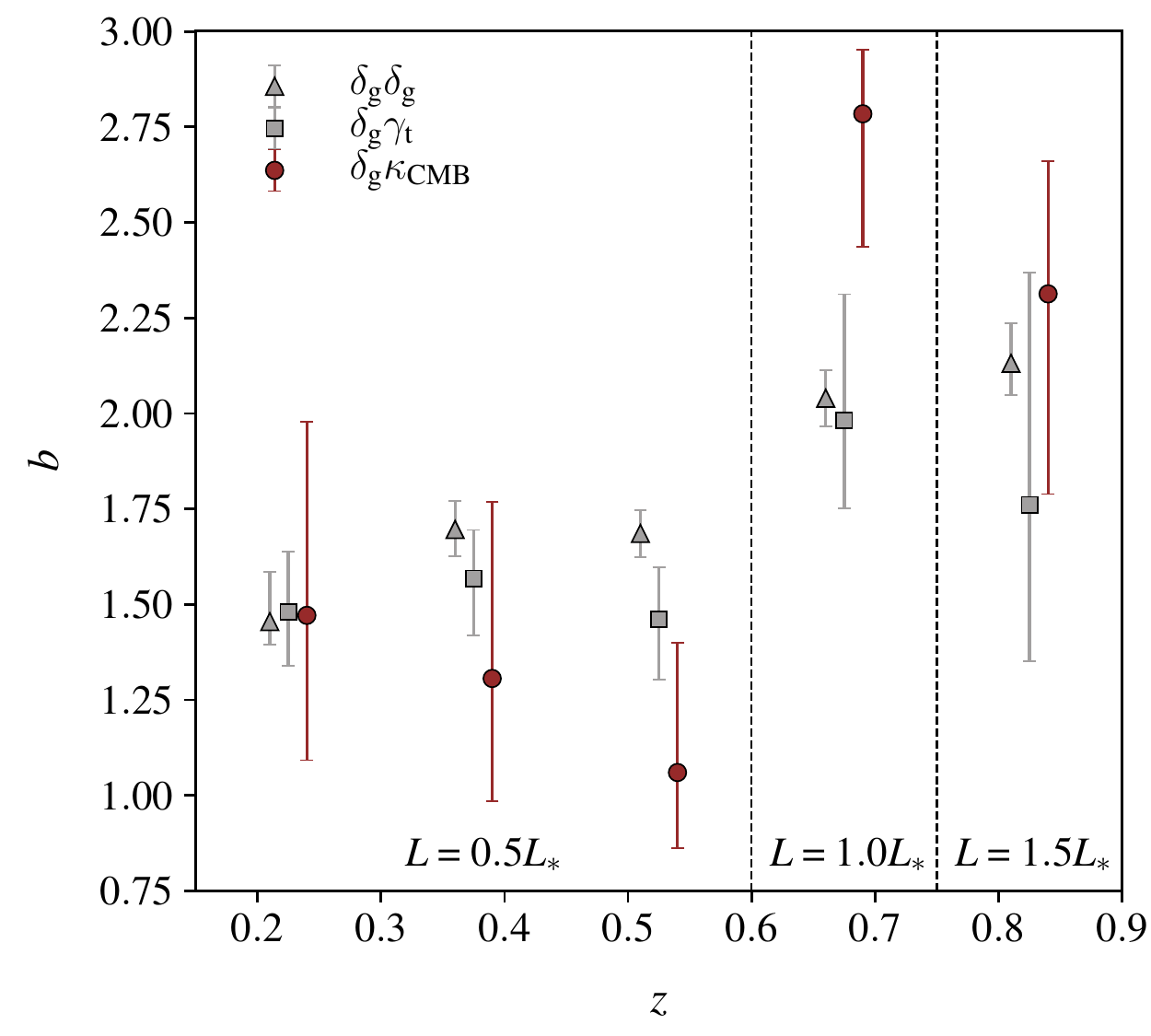}
  \caption{Galaxy bias estimation of \redmagic\, galaxies from different probes, as a function of redshift
when combined with $\planck$ baseline likelihood.
The results {from $w^{\delg\kcmb}(\theta)$} assuming $A_{\kappa}=1$ are shown in red. Additionally plotted are the measurements from galaxy clustering \citep{ElvinPoole2017} (gray triangles) and galaxy-galaxy lensing \citep{Prat2017} (gray squares). 
 }
\label{fig:bias}
\end{center}
\end{figure}

\subsection{Growth constraints}\label{sec:implications}
Next, we study the broader cosmological implications of our measurement: we first assume cosmology to be fixed at high redshift, e.g. by the $\planck$ CMB observations, and test whether the linear growth function inferred from our measurement at low redshift is consistent with the baseline \LCDM model predictions. We test this using the two methods described in Sec.~\ref{sec:Dg_theory}.

\subsubsection{The $D_{\rm G}$ estimator}
\label{sec:Dg}

We first measure the linear growth function $D(z)$ using the $D_{\rm G}$ estimator. We compute $D_{\rm G}$ for all five tomographic bins, applying the conservative angular scale cuts listed in Sec.~\ref{sec:scalecuts}, and additionally removing scales above $100'$ (see Sec.~\ref{sec:Dg_theory} for details).
The results are shown in Fig.~\ref{fig:EG}.
In addition, we also calculate the best-fit amplitude $A_{\rm D}$ by combining all the bins, from which we obtain $A_{\rm D} = 1.16^{+0.20}_{-0.20}$, which agrees with the fiducial \LCDM expectation of $A_{\rm D}=1$.

In comparing the result presented here (using 1289 $\sqdeg$) with that of \citetalias{Giannantonio2016} that used DES-SV data covering 140 $\sqdeg$, we see that we obtain similar {constraining power}. This is due to:
\begin{enumerate}
\item The conservative scale cuts imposed in this study to minimize $w^{\delg\ktsz}(\theta)$ bias.
\item The lower number density of $\redmagic$ than the galaxy sample used by \citetalias{Giannantonio2016}, in exchange for better photo-$z$ errors.
\item The lower temperature $\ell_{\rm max}$ used ($\ell_{\rm max}=3000$) in the lensing reconstruction process in \cite{Omori2017} to avoid contamination by astrophysical foregrounds
(whereas the SPT lensing map used in the \citetalias{Giannantonio2016} analysis had $\ell_{\rm max}=4000$).
\end{enumerate}
Therefore, we have exchanged signal-to-noise ratio with increased robustness of the measurement.

We have tested that the general scale cuts used by \cite{Baxter2018} are also appropriate for the $D_{\rm G}$ estimator. We have confirmed this by running the estimator on contaminated theory data vectors, for which we found that the bias on the recovered growth is always $< 0.5 \sigma$ if the standard scale cuts of Sec.~\ref{sec:scalecuts} are used. This is not the case for less conservative cuts, which we therefore discard: for example, using the full range of scales down to $2'$ biases $D_{\rm G}$ at the $2 \sigma$ level. {The bias at small angular scales is mainly driven by the tSZ contamination in the CMB lensing map}, {as} discussed in \cite{Baxter2018}.

\begin{figure}
\begin{center}
\includegraphics[width=1.0\linewidth, angle=0]{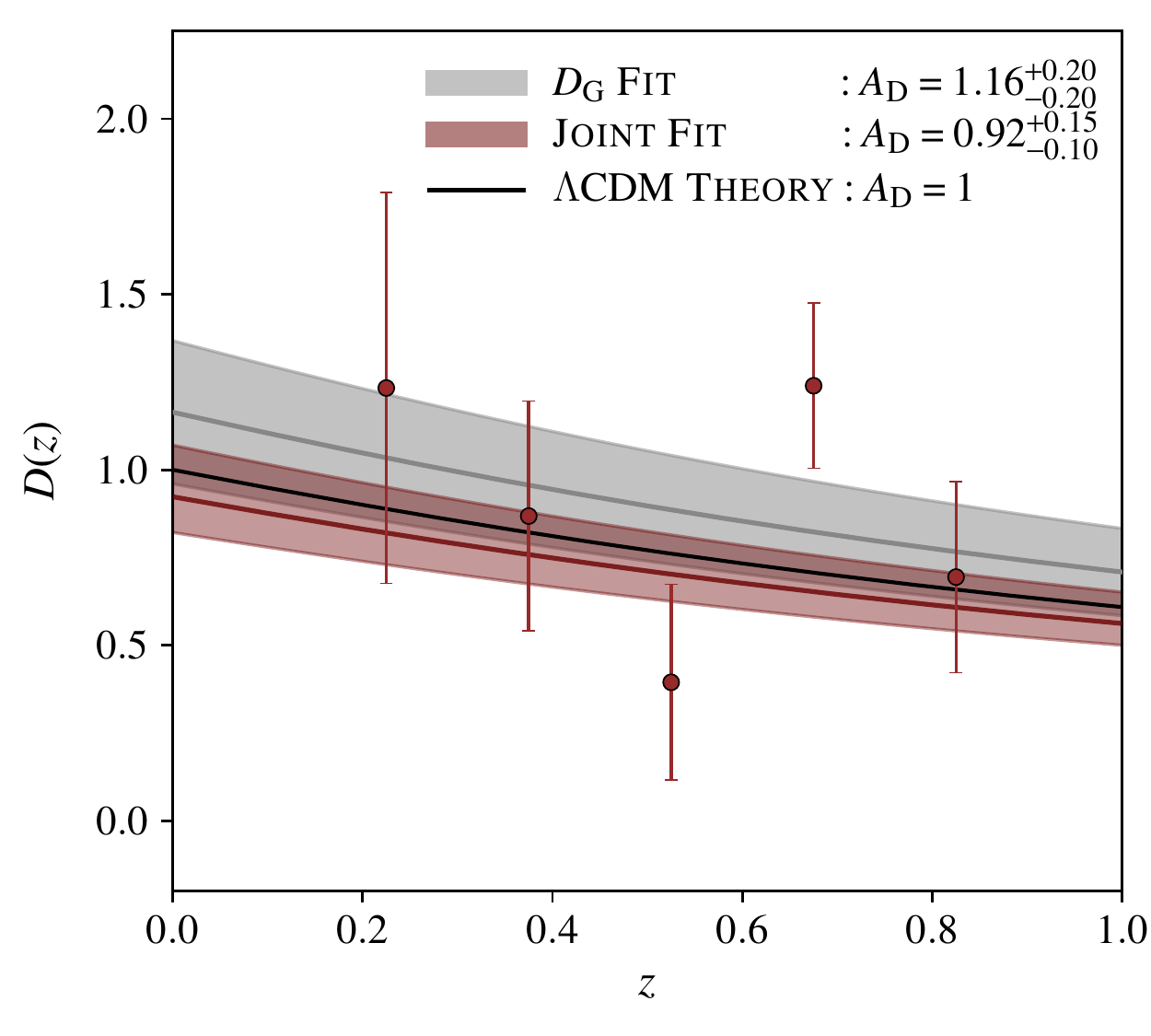}
  \caption{Growth function estimates from the combination of auto- and cross-correlation functions, at the fiducial cosmology. The red points show the measured value of $D_{\rm G}$ in each redshift bin, with error bars representing the diagonal elements of the covariance matrix described in Sec.~\ref{sec:Dg_theory}. The grey band represents the $1\sigma$ confidence interval on the best-fit amplitude $A_{\rm D}$, assuming the fiducial $\Lambda$CDM template shown in black (solid), and the red shaded regions describe the 1-$\sigma$ uncertainties from the {joint-fit} analysis described in Sec.~\ref{sec:MCMC_Grow}.}
\label{fig:EG}
\end{center}
\end{figure}

\subsubsection{Joint growth fit results}
\label{sec:MCMC_Grow}

Here we keep the cosmological parameters fixed to the fiducial model, but {marginalize} over the five independent linear galaxy bias parameters (one for each redshift bin), the photo-$z$ uncertainties and the linear growth parameter $A_{\rm D}$ using the priors presented in Table~\ref{tab:priors}. We measure the linear galaxy bias to be $b_1 = 1.45^{+0.30}_{-0.15}$, $b_2 = 1.73^{+0.26}_{-0.22}$, $b_3 = 1.80^{+0.17}_{-0.29}$, $b_4 = 2.04^{+0.35}_{-0.21}$, $b_5 = 2.15^{+0.36}_{-0.24}$ and find a constraint of $A_{\rm D} = 0.92^{+0.15}_{-0.10}$  for the amplitude of {the growth function}.  These measurements of the bias are in agreement with the results shown in Table \ref{tab:fitresults}. The recovered growth function agrees with the fiducial \LCDM expectation, as the measurement of $A_{\rm D}$ is consistent with 1.0. {We observe that the} errors on the {galaxy} bias are larger compared to a direct best-fit estimation {presented in Sec.~\ref{sec:directfit}}. This is due to the fact that the bias and $A_{\rm D}$ parameters are correlated.  In turn, the fact that the joint-fit method gets a different value of  $A_{\rm D}$ with respect to the $D_{\rm G}$ method is because it explicitly takes into account the correlations between bias and growth.

\subsection{Cosmological parameter estimation}
\label{sec:MCMC_Full}

In this section, we present the full cosmological analysis using the $w^{\delg\delg}(\theta)$+$w^{\delg\kcmb}(\theta)$ data vectors  and marginalizing over all the cosmological parameters and nuisance parameters (galaxy bias and photo-$z$ bias, but we fix $A_{\rm D}=A_{\kappa}=1$).

The priors used in this analysis are summarized in Table~\ref{tab:priors}, and are {the same as used in}  \cite{DES2017cosmo} to maintain consistency between the analyses.

In Fig.~\ref{fig:Om_S8} we show the constraints obtained on matter density $\omm$ and 
$S_8$ {when all six cosmological parameters, photo-$z$ errors and linear galaxy biases for the five redshift bins are marginalized over}.
This is then compared with the constraints from the combination of $w^{\delg\delg}(\theta)+w^{\delg\gammat}(\theta)$ as presented in \cite{DES2017cosmo}. We observe that these two measurements slice through the parameter space slightly differently.  Using $w^{\delg\delg}(\theta)+w^{\delg\kcmb}(\theta)$ we obtain $\Omega_{\rm m}=0.276^{+0.029}_{-0.030}$ and $S_8=0.800^{+0.090}_{-0.094}$, whereas the combination of $w^{\delg\delg}(\theta)+w^{\delg\gammat}(\theta)$ gives us $\Omega_{\rm m}=0.294^{+0.047}_{-0.029}$ and $S_8=0.759^{+0.037}_{-0.031}$. These two results can also be compared with the constraints from the combination of $w^{\delg\delg}(\theta)+w^{\delg\gammat}(\theta)+\xi_{+/-}(\theta)$ (also referred to as $3\times2$pt \citep{DES2017cosmo}), which gives $\Omega_{\rm m}=0.267^{+0.030}_{-0.017}$, $S_{8}=0.773^{+0.026}_{-0.020}$. These results are highly consistent with each other as shown on Fig. \ref{fig:Om_S8}.

The measurement used in this analysis are combined with the $w^{\gammat\kcmb}(\theta)$ presented in \citep{DESY1_shearkcmb} and the results from \citep{DES2017cosmo} in \citep{DESY1_5x2}, using the methodology outlined in \cite{Baxter2018}.

\begin{figure}
\begin{center}
\includegraphics[width=1.00\linewidth]{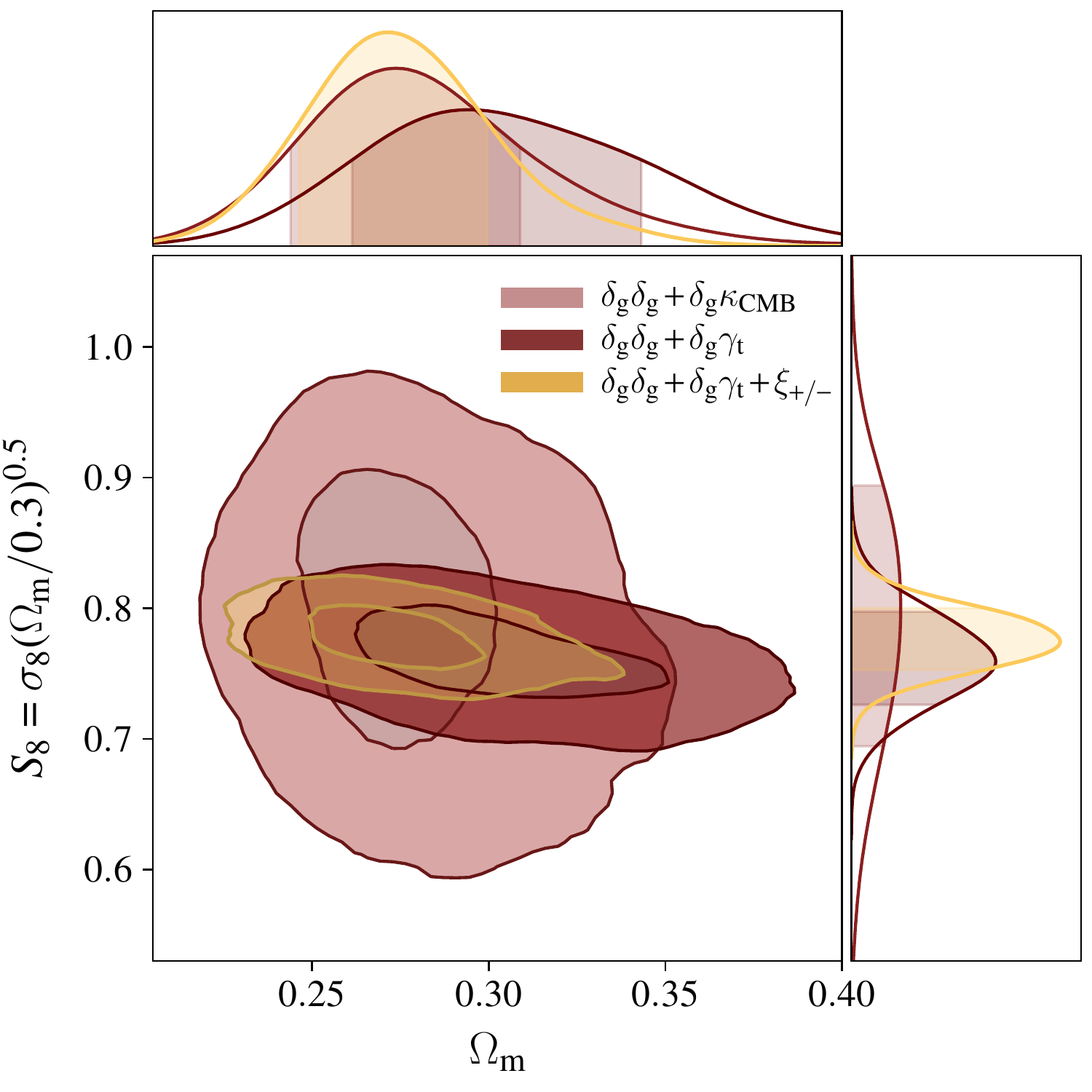}
\caption{Constraints on $\omm$ and $S_{8}$ from the measurements of this paper and combinations of other two-point correlation functions. We show the constraints from $w^{\delg\delg}(\theta)+w^{\delg\kcmb}(\theta)$ (light red), $w^{\delg\delg}(\theta)+w^{\delg\gammat}(\theta)$ (dark red) and $w^{\delg\delg}(\theta)+w^{\delg\gammat}(\theta)+\xi_{+/-}(\theta)$ (gold), where $\xi_{+/-}(\theta)$ are the cosmic shear measurements. While the constraints obtained from $w^{\delg\delg}(\theta)+w^{\delg\kcmb}(\theta)$ are weaker than those of $w^{\delg\delg}(\theta)+w^{\delg\gammat}(\theta)$, they are  consistent with those obtained from $w^{\delg\delg}(\theta)+w^{\delg\gammat}(\theta)+\xi_{+/-}(\theta)$, and have slightly different degeneracy directions.}
\label{fig:Om_S8}
\end{center}
\end{figure}

\section {Conclusions} \label{sec:conclusion}
We have presented measurements of the DES  $\redmagic$ galaxy-CMB lensing cross-correlation as a function of redshift.
Our measurement rejects the hypothesis of no-lensing at 19.9$\sigma$\footnote{We note that while certain systematics could add to the apparent signal and artificially inflate the significance, in this case the main contamination without scale cuts is tSZ, which artificially reduces $w^{\delg\kcmb}$. In other words, in the absence of tSZ and scale cuts, the significance of this measurement would be higher than 19.9 $\sigma$.} significance prior to any scale cuts and 9.9$\sigma$ using the conservative scale cuts from \cite{Baxter2018}. The conservative scale cuts reduce the signal to noise of the measurements in exchange for mitigation of systematic biases.

We test for the impact of possible systematics in the cross-correlations, considering contaminants to both the DES galaxy and the CMB lensing maps. We find that, on the scales we consider, all these contaminants have a small impact on our measurements compared with the statistical uncertainties. The largest effect comes from the tSZ contribution to the CMB lensing maps, which becomes large at smaller angular scales, and is the main limiting factor dictating our scale cuts \citep{Baxter2018}. Improving the modeling and subtraction of this contaminant will be the key to extracting the full statistical power of the temperature based CMB lensing maps in the future.

In obtaining the galaxy bias parameters, we find that galaxy-clustering measurements place significantly tighter constraints than galaxy-galaxy lensing or galaxy-CMB lensing correlations. However, the two cross-correlations are nonetheless important in breaking degeneracies in parameter space. 

We use our measurements to infer cosmological information in a number of ways. We first constrain the linear growth function using the $D_{\rm G}$ estimator introduced by \citetalias{Giannantonio2016}, finding a relative growth amplitude of $A_{\rm D} = 1.16^{+0.20}_{-0.20}$. This compiles measurements of growth in various tomographic bins, accounting for their covariance.
We then extend this result and constrain the relative growth amplitude with a joint-fit method, marginalizing over galaxy biases and photo-$z$ uncertainties, and considering the full covariance of the observables. In this case, we find $A_{\rm D} = 0.92 ^{+0.15}_{-0.10}$. Both of these results are consistent with the $\Lambda {\rm CDM}$ predictions of $A_{\rm D}=1$.

Using these measurements, we finally run a full MCMC analysis over the \LCDM cosmological parameters to also place marginalized constraints on the two parameters that are most directly related to the matter density field: $\Omega_{\rm m}$ and $S_8 \equiv \sigma_{8}\sqrt{\Omega_{\rm m}/0.3}$. Using the combination of $w^{\delg\delg}(\theta)$ and $w^{\delg\kcmb}(\theta)$ we obtain {$\Omega_{\rm m}=0.276^{+0.029}_{-0.030}$} and {$S_8=0.800^{+0.090}_{-0.094}$}.  This can be compared with the results obtained using galaxy clustering and galaxy-galaxy lensing (i.e. $w^{\delg\delg}(\theta)$+$w^{\delg\gammat}(\theta)$), which gives $\Omega_{\rm m}=0.294^{+0.047}_{-0.029}$ and $S_8=0.759^{+0.037}_{-0.031}$. While our measurement here is less constraining in $S_8$, we are able to obtain a tighter constraint on $\Omega_{\rm m}$: the extra redshift bin at $z\sim1089$ that the CMB lensing information provides improves the constraints. As can be inferred from Fig. \ref{fig:Om_S8}, the cosmological information that we can extract from the combination of $w^{\delg\delg}(\theta)$ and $w^{\delg\kcmb}(\theta)$ correlations is not completely degenerate with the information that could be extracted from $w^{\delg\delg}(\theta)$ and $w^{\delg\gammat}(\theta)$. Furthermore, we have found that the constraining power is comparable despite the conservative scale cuts we have applied in this analysis, and we expect to obtain better signal in the future as we use galaxy samples at higher redshifts due to the better overlap with the CMB lensing kernel. 

The constraining power of DES measurements of galaxy-CMB lensing correlations has the potential to improve in future analyses. The DES-Y3 will cover the full 5000 $\sqdeg$ of the DES footprint at approximately the same depth as Y1. Since the extended area does not overlap with the SPT footprint, we expect the gain in the signal-to-noise to be small in terms of improvements in sky coverage. However, our analysis choice in this study is conservative; we have chosen the scale-cuts to minimize the biases in exchange for signal-to-noise ratio. To improve this measurement further, it will be essential to (1) characterize the bias to a higher accuracy, such that the signal loss is minimized, or (2) to improve the reconstruction of the CMB lensing map so that it is less prone to biases \citep[see e.g.][for a discussion of modifications to temperature-based lensing reconstruction to minimize tSZ bias]{Madhavacheril2018}.  Furthermore, newer data sets from SPT (SPTpol and SPT-3G \citep{Benson14}) have lower noise levels than SPT-SZ, and therefore, lensing maps generated from these data sets will have lower noise. Improvements along these lines will allow us to maximally extract the signal from this cross-correlation, and to reach the best possible accuracy on cosmology.

\section*{Acknowledgements}

This paper has gone through internal review by the DES collaboration.

YO acknowledges funding from the Kavli Foundation, the Natural Sciences and Engineering Research Council of Canada, Canadian Institute for Advanced Research, and Canada Research Chairs program. EB is partially supported by the US Department of Energy grant DE-SC0007901. CC was supported in part by the Kavli Institute for Cosmological Physics at the University of Chicago through grant NSF PHY-1125897 and an endowment from Kavli Foundation and its founder Fred Kavli. CR acknowledges support from a Australian Research Council Future Fellowship (FT150100074). BB is supported by the Fermi Research Alliance, LLC under Contract No. De-AC02-07CH11359 with the United States Department of Energy.
The McGill authors acknowledge funding from the Natural Sciences and Engineering Research Council of Canada, Canadian Institute for Advanced Research, and Canada Research Chairs program.

The South Pole Telescope program is supported by the National Science Foundation through grant PLR-1248097. Partial support is also provided by the NSF Physics Frontier Center grant PHY-0114422 to the Kavli Institute of Cosmological Physics at the University of Chicago, the Kavli Foundation, and the Gordon and Betty Moore Foundation through Grant GBMF\#947 to the University of Chicago.
This publication makes use of data products from the Wide-field Infrared Survey Explorer, which is a joint project of the University of California, Los Angeles, and the Jet Propulsion Laboratory/California Institute of Technology, funded by the National Aeronautics and Space Administration.
Argonne National Laboratory's work was supported under U.S. Department of Energy contract DE-AC02-06CH11357.

Funding for the DES Projects has been provided by the U.S. Department of Energy, the U.S. National Science Foundation, the Ministry of Science and Education of Spain, 
the Science and Technology Facilities Council of the United Kingdom, the Higher Education Funding Council for England, the National Center for Supercomputing 
Applications at the University of Illinois at Urbana-Champaign, the Kavli Institute of Cosmological Physics at the University of Chicago, 
the Center for Cosmology and Astro-Particle Physics at the Ohio State University,
the Mitchell Institute for Fundamental Physics and Astronomy at Texas A\&M University, Financiadora de Estudos e Projetos, 
Funda{\c c}{\~a}o Carlos Chagas Filho de Amparo {\`a} Pesquisa do Estado do Rio de Janeiro, Conselho Nacional de Desenvolvimento Cient{\'i}fico e Tecnol{\'o}gico and 
the Minist{\'e}rio da Ci{\^e}ncia, Tecnologia e Inova{\c c}{\~a}o, the Deutsche Forschungsgemeinschaft and the Collaborating Institutions in the Dark Energy Survey. 

The Collaborating Institutions are Argonne National Laboratory, the University of California at Santa Cruz, the University of Cambridge, Centro de Investigaciones Energ{\'e}ticas, 
Medioambientales y Tecnol{\'o}gicas-Madrid, the University of Chicago, University College London, the DES-Brazil Consortium, the University of Edinburgh, 
the Eidgen{\"o}ssische Technische Hochschule (ETH) Z{\"u}rich, 
Fermi National Accelerator Laboratory, the University of Illinois at Urbana-Champaign, the Institut de Ci{\`e}ncies de l'Espai (IEEC/CSIC), 
the Institut de F{\'i}sica d'Altes Energies, Lawrence Berkeley National Laboratory, the Ludwig-Maximilians Universit{\"a}t M{\"u}nchen and the associated Excellence Cluster Universe, 
the University of Michigan, the National Optical Astronomy Observatory, the University of Nottingham, The Ohio State University, the University of Pennsylvania, the University of Portsmouth, 
SLAC National Accelerator Laboratory, Stanford University, the University of Sussex, Texas A\&M University, and the OzDES Membership Consortium.

Based in part on observations at Cerro Tololo Inter-American Observatory, National Optical Astronomy Observatory, which is operated by the Association of 
Universities for Research in Astronomy (AURA) under a cooperative agreement with the National Science Foundation.

The DES data management system is supported by the National Science Foundation under Grant Numbers AST-1138766 and AST-1536171.
The DES participants from Spanish institutions are partially supported by MINECO under grants AYA2015-71825, ESP2015-66861, FPA2015-68048, SEV-2016-0588, SEV-2016-0597, and MDM-2015-0509, 
some of which include ERDF funds from the European Union. IFAE is partially funded by the CERCA program of the Generalitat de Catalunya.
Research leading to these results has received funding from the European Research
Council under the European Union's Seventh Framework Program (FP7/2007-2013) including ERC grant agreements 240672, 291329, and 306478.
We  acknowledge support from the Australian Research Council Centre of Excellence for All-sky Astrophysics (CAASTRO), through project number CE110001020, and the Brazilian Instituto Nacional de Ci\^encia
e Tecnologia (INCT) e-Universe (CNPq grant 465376/2014-2).

This manuscript has been authored by Fermi Research Alliance, LLC under Contract No. DE-AC02-07CH11359 with the U.S. Department of Energy, Office of Science, Office of High Energy Physics. The United States Government retains and the publisher, by accepting the article for publication, acknowledges that the United States Government retains a non-exclusive, paid-up, irrevocable, world-wide license to publish or reproduce the published form of this manuscript, or allow others to do so, for United States Government purposes.
Computations were made on the supercomputer Guillimin from McGill University, managed by Calcul Qu\'{e}bec and Compute Canada. The operation of this supercomputer is funded by the Canada Foundation for Innovation (CFI), the minist\`{e}re de l'\'{E}conomie, de la science et de l'innovation du Qu\'{e}bec (MESI) and the Fonds de recherche du Qu\'{e}bec - Nature et technologies (FRQ-NT).
This research is part of the Blue Waters sustained-petascale computing project, which is supported by the National Science Foundation (awards OCI-0725070 and ACI-1238993) and the state of Illinois. Blue Waters is a joint effort of the University of Illinois at Urbana-Champaign and its National Center for Supercomputing Applications.

We acknowledge the use of many python packages: \textsc{Astropy}, a community-developed core Python package for Astronomy \citep{astropy18}, \textsc{CAMB} \citep{Lewis00,Howlett12}, \textsc{Chain consumer}\footnote{\url{https://samreay.github.io/ChainConsumer}}, \textsc{CosmoSIS}\footnote{\url{https://bitbucket.org/joezuntz/cosmosis}},
\textsc{HEALPix} \citep{gorski05}, \textsc{IPython} \citep{ipython07}, \textsc{Matplotlib} \citep{hunter07}, \textsc{NumPy} \& \textsc{SciPy} \citep{scipy,numpy}, \textsc{Quicklens}\footnote{\url{https://github.com/dhanson/quicklens}} and \textsc{TreeCorr} \citep{jarvis15}.

\bibliography{ms}

\appendix

\section{Covariance matrix validation}
\label{sec:covmatJK}

Here we compare the analytical covariance matrix developed in \cite{Krause2017} (and extended to include $\kcmb$ in \cite{Baxter2018}) with the data-based covariance obtained using the jackknife method. 

From the data we compute the jackknife covariance as
\begin{equation}
C^{\rm jk}_{ij}=\frac{N_{\rm jk}-1}{N_{\rm jk}}\sum_{k}^{N_{\rm jk}}(\vec{d}^{k}-\bar{d})_{i}(\vec{d}^{k}-\bar{d})_{j}, 
\end{equation}
where $N_{\rm jk}$ is the number  of jackknife patches used. Fig.~\ref{fig:fig_jkdiag} shows the comparison between the jackknife and analytical covariances.
We {obtain} diagonal covariance elements {that are} on average $\sim 17\%$ higher than the analytical covariance over the angular scales of interest. Based on the discussion in \cite{Norberg2009} and \cite{Friedrich2016}, we consider that these are in sufficient agreement given that the jackknife method is a noisy estimate of the underlying covariance.

\begin{figure}
\begin{center}
\includegraphics[width=0.48\textwidth]{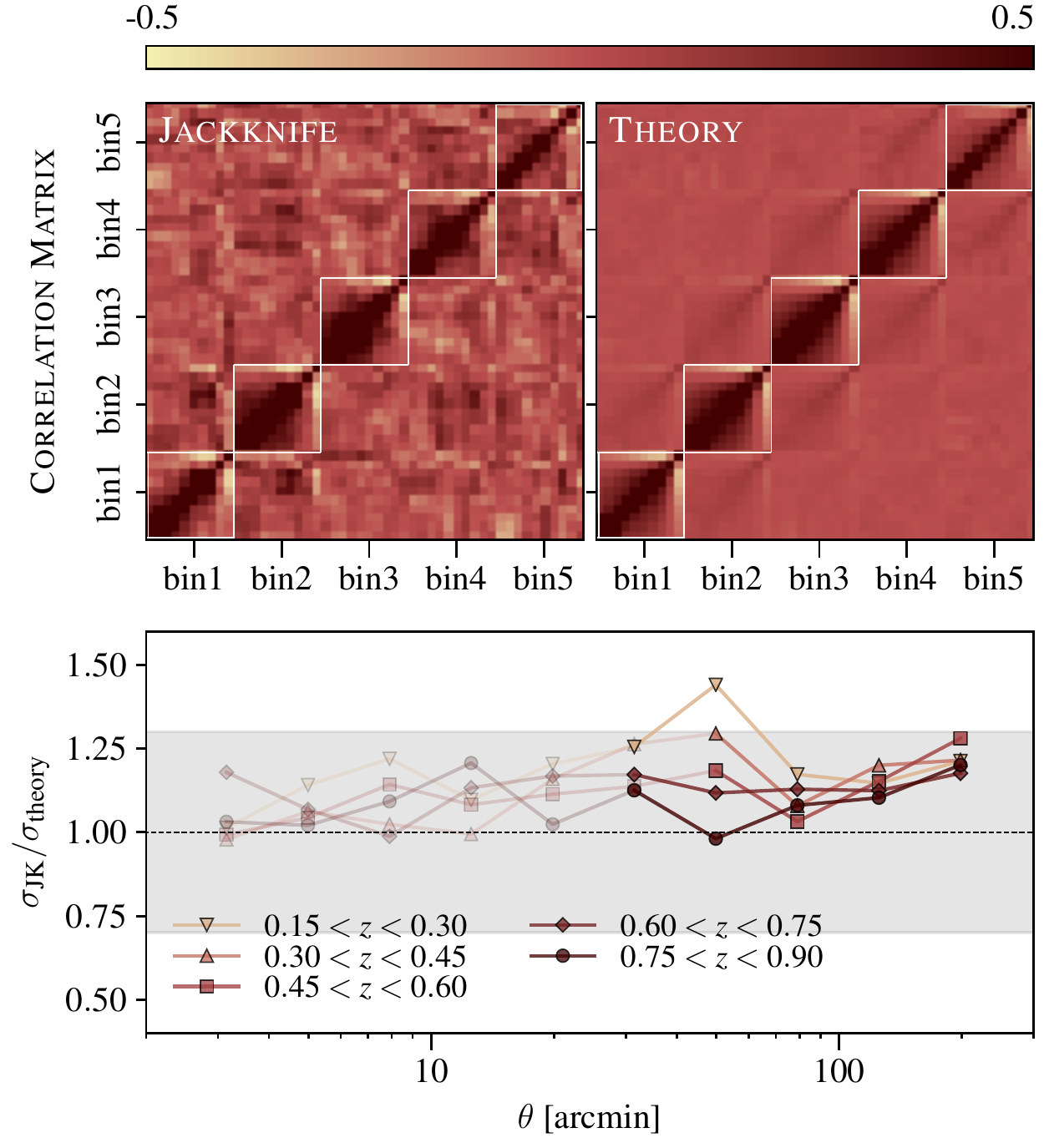}
\caption{Covariance matrix for $w^{\delg\kcmb}(\theta)$ obtained using the jackknife method (upper left) and analytically (upper right). Each redshift bin includes 10 angular bins.  The ratio of the diagonal elements are shown in the lower panel. The gray panel denotes $\pm30\%$ margin compared to the analytical covariance. }
\label{fig:fig_jkdiag}
\end{center}
\end{figure}

\label{lastpage}

\end{document}